\documentclass{aa} 
\usepackage{graphicx}
\usepackage{txfonts}
\usepackage{multirow}
\usepackage{caption, threeparttable}
\usepackage{xcolor}
\usepackage{float}
\usepackage{amsmath}
\usepackage[colorlinks=true,linkcolor=blue, citecolor=blue]{hyperref}%

\begin{document}

\newcommand{\teff}{$T_\mathrm{eff}$}
\newcommand{\logg}{$\log\,g$}
\newcommand{\feh}{[M/H]}
\newcommand{\microturb}{$\xi_\mathrm{micro}$}
\newcommand{\sife}{[Si/Fe]}
\newcommand{\mgfe}{[Mg/Fe]}
\newcommand{\afe}{[$\alpha$/Fe]}
\newcommand{\co}{CO($\nu =2-0$)\,}

\newcommand{\water}{H$_2$O}
\newcommand{\invcm}{cm$^{-1}$}
\newcommand{\kms}{km\,s$^{-1}$}
\newcommand{\mic}{$\mu \mathrm m$}
\newcommand{\msun}{$\mathrm{M}_\odot$}

\newcommand{\GKcomment}[1]{\textcolor{red}{(GK: \it #1)}}
\newcommand{\GKchange}[1]{\textcolor{red}{\bf #1}}
\newcommand{\Mathias}[1]{{\color{green} [Mathias:#1]}}
\newcommand{\GNcomment}[1]{\textcolor{brown}{(GN: \it #1)}}
\newcommand{\PAPcomment}[1]{\textcolor{purple}{(PAP: \it #1)}}

\defcitealias{Nandakumar:2023_I}{Paper\,I}
\defcitealias{Nandakumar:2023b_II}{Paper\,II}
\defcitealias{Nandakumar:24_III}{Paper\,III}
\titlerunning{Massive stars in Gaia DR3} 
\authorrunning{Nandakumar et al.}

   \title{Reassessing the low $\alpha$ massive sequence stars in Gaia RVS}

   \author{
            G. Nandakumar
            \inst{1,2}
            \and
           G. Kordopatis
          \inst{3}
          \and
           M. Schultheis
          \inst{3}
          \and
           N. Ryde
          \inst{2}
          \and
           P. A. Palicio 
          \inst{3,8,9}
          \and
          E. Spitoni
          \inst{4,5}
          \and
          F. Matteucci
          \inst{4,6,7}
          }

   \institute{
   Aryabhatta Research Institute of Observational Sciences, Manora Peak, Nainital 263001, India\\
              \email{tbd}
            \and
            Division of Astrophysics, Department of Physics, Lund University, Box 118, SE-221 00 Lund, Sweden
            \and 
            Universit\'e C\^ote d’Azur, Observatoire de la C\^ote d’Azur, CNRS, Laboratoire Lagrange, Nice, France\label{oca}
            \and
            INAF – Osservatorio Astronomico di Trieste, via G.B. Tiepolo 11, 34143 Trieste, Italy
            \and
            IFPU, Institute for Fundamental Physics of the Universe, Via Beirut 2, 34151 Trieste, Italy
            \and
            Dipartimento di Fisica, Sezione di Astronomia, Università di Trieste, Via G. B. Tiepolo 11, 34143 Trieste, Italy
            \and
            INFN Sezione di Trieste, via Valerio 2, 34134 Trieste, Italy
            \and
            Instituto de Astrof\'isica de Canarias, E-38205 La Laguna, Tenerife, Spain
            \and
            Universidad de La Laguna, Dpto. Astrof\'isica, E-38206 La Laguna, Tenerife, Spain
 }

   \date{Received ; accepted }

\abstract
{Studying the chemical abundance of stars of different ages and masses allows us to trace the chemical evolution of the Milky Way. Recently, a chemically depleted young massive stellar population was identified using the spectroscopic catalogue of Gaia DR3. To explain its characteristics, a recent enhanced star formation event, via a third infall occurring within the last 2 Gyr, has been evoked. }
   {In this paper we reassess the low $\alpha$-sequence of massive stars identified in the \textit{Gaia} spectroscopic catalog and investigate their presence in other Milky Way spectroscopic survey catalogs.}  
   { We select massive sequence stars and red giant branch stars from the \textit{Gaia} DR3 catalogue using the same filtering strategy adopted in previous chemical-cartography studies. These samples are then cross-matched with APOGEE DR17, GALAH DR4, and Gaia-CNN to enable a detailed comparison of stellar parameters and $\alpha$-abundances. Stellar masses are estimated by projecting their atmospheric parameters and infrared magnitudes onto PARSEC isochrones.}
   { For the massive-star sample, we find large discrepancies in stellar parameters and calcium abundances between Gaia DR3 and the three external surveys. The external catalogues do not show a low-calcium sequence but rather resemble those of thin-disc RGB stars. Other $\alpha$-elements (silicon in APOGEE and GALAH, and magnesium in GALAH) also do not show depleted values. In APOGEE, however, massive sequence stars with metallicities above –0.5\,dex display lower magnesium abundances. We attribute this to APOGEE’s use of macroturbulence velocities calibrated solely on metallicity. }
{Our analysis does not show any evidence for $\alpha$-element depletion in massive sequence stars. Alpha-abundances of massive sequence stars derived from the Gaia-RVS spectra should therefore be used with  caution. Nevertheless, the previously proposed three-infall chemical-evolution models remain plausible: even without a chemically depleted young massive population, scenarios involving only mild dilution could still account for recent star-formation episodes.  }


   \maketitle
%

\section{Introduction}
\label{sec:intro}
\vspace{-5pt}

Over the last decade, the \textit{Gaia} mission \citep{Gaia:2016} has been instrumental in furthering our understanding about the Milky Way structure and its various components by providing very precise astrometric, parallax, and proper motion measurements of the Milky Way stellar populations. The stellar orbits and kinematic properties of stars determined from the \textit{Gaia} data have made possible the discovery of several substructures/merger debris in the Milky Way halo and disc \citep[see, e.g.,][]{Bonaca:2017,Helmi:2018,Belokurov:2018,malhan:2018,dimatteo:2019,myeong:2019,Belokurov:2020}. The detailed chemo-dynamic investigations of these substructures resulting from combining the kinematic information from  \textit{Gaia} with the elemental abundances from high-resolution spectroscopic surveys of the Milky Way such as Apache Point Observatory Galactic Evolution Experiment/APOGEE \citep{Majewski:2017}, Galactic archaeology with HERMES/GALAH \citep{DeSilva:2015} or Gaia-ESO survey \citep{Gilmore:2012} have led to several hypotheses on the merger history of the Milky Way \citep[see the review of][and references therein]{Deason:24}. 

The latest Gaia data release \citep[DR3;][]{GDR3:2023} has made public the atmospheric parameters (\teff, \logg, \feh) of $\sim$5.6 million sources analysed by the General Stellar Parametriser - spectroscopy \citep[GSP-spec,][]{Recio-Blanco:2023} module. This module uses the medium resolution (R$\sim$11,500) spectra collected by the on-board Radial Velocity Spectrometer \citep[RVS;][]{Cropper:2018} as input, and additionally provides few elemental abundances for a subset of stars \citep{Fouesneau:2023}\footnote{hereafter, we use the terminology \textit{Gaia} DR3 when referring to the GSP-Spec output.} . With this chemical information, \cite{Recio-Blanco:2023chemcart} carried out a detailed chemical cartography of the Milky Way based on chemo-dynamical analysis of the disc and halo populations. In addition to several findings about the disc structure morphology, chemical patterns associated with the kinematic or orbital substructures in the disc, radial metallicity gradient-age relation with the samples of open clusters, their study also revealed chemical impoverishment in calcium in the metallicity range of -0.5 to +0.0\,dex for young and massive disc stars. These stars were chosen based on  cuts in \teff\, and uncalibrated \logg\,(see their Figure 8). They also reported low cerium abundances, [Ce/Fe], in the coolest part of the massive stars with low calcium stars located at the position of the Galactic arms (see their Figure 12). They noted that this observed chemical depletion cannot be explained by stellar migration since the decreasing stellar density profile with the galactocentric distance does not favour enough inwards stellar migration that would decrease the local metallicities and lead to such a young metal-impoverished stellar population. \cite{Spitoni:2023} proposed a three-infall chemical evolution model that could reproduce the observed low alpha abundances. The proposed model is constrained by several recent (last 2 Gyr) enhanced star formation episodes consistent with the Sagittarius dwarf spheroidal galaxy pericentre passages as inferred by \cite{Ruiz-Lara:2020} from Gaia DR2 colour–magnitude diagrams. \cite{Spitoni:2023} further showed that a subset of this young population observed in APOGEE have low magnesium abundances (see their Figure 14). But there have been no investigation on whether other alpha element species in APOGEE, especially calcium,  also show similar depletion. It is also pertinent to confirm the chemical depletion in the common stars observed in other dedicated high resolution spectroscopic surveys such as GALAH. 

In this work, we investigate the chemically depleted massive stars (hereafter named massive sequence stars\footnote{The Gaia RVS paper mistakenly calls them massive stars, but since this term refers to the stars that can go through core-collapse, we prefer to call these stars massive sequence stars to avoid confusion in potential future follow-ups.}) identified in the \textit{Gaia} DR3 catalog by detailed comparison of the stellar parameters and alpha abundances of common stars in other spectroscopic survey catalogs. For this, we choose APOGEE, GALAH, and \textit{Gaia}-CNN as the external catalogs, and cross match them with the massive and RGB stars selected from the \textit{Gaia} DR3 catalog applying the selection criteria used in \cite{Recio-Blanco:2023chemcart}. We choose APOGEE and GALAH due to their high resolution (R$\sim$22,000 and 28,000 respectively) and dedicated analysis pipelines, that thus should provide reliable stellar parameters and abundances. With \textit{Gaia} CNN, we could compare the stellar parameters and abundances estimated from the same spectra used to produce the \textit{Gaia} DR3 catalog but applying a different method and with additional information. RGB stars did not show any chemical depletion and were used as a comparison sample in \cite{Recio-Blanco:2023chemcart}, which we also follow in this work. For these selected stars, we also computed the masses independently by projecting the stellar parameters from each catalog and their 2MASS \citep{2mass} and WISE \citep{WISE} photometric magnitudes on the PARSEC isochrone models \citep{Bressan2012}. With these data sets, we aim to reassess the low $\alpha$ massive sequence stars identified in the \textit{Gaia} DR3. 

Brief descriptions of the data sets used in this work are given in Section~\ref{sec:dataset}, followed by the details of the applied selection cuts to select massive and RGB stars in Section~\ref{sec:selection}. The mass computation method and validation of the computed masses are described in Section~\ref{sec:masses}. In Section~\ref{sec:comparison}, we compare the stellar parameters, alpha abundances and masses of the common massive and RGB stars in \textit{Gaia} DR3 and the three external catalogs. Finally, we discuss the results from our investigation in Section~\ref{sec:discussion}, followed by the conclusions in Section~\ref{sec:conclusions}. 

 \begin{figure*}
 \centering
  \includegraphics[width=0.8\textwidth]{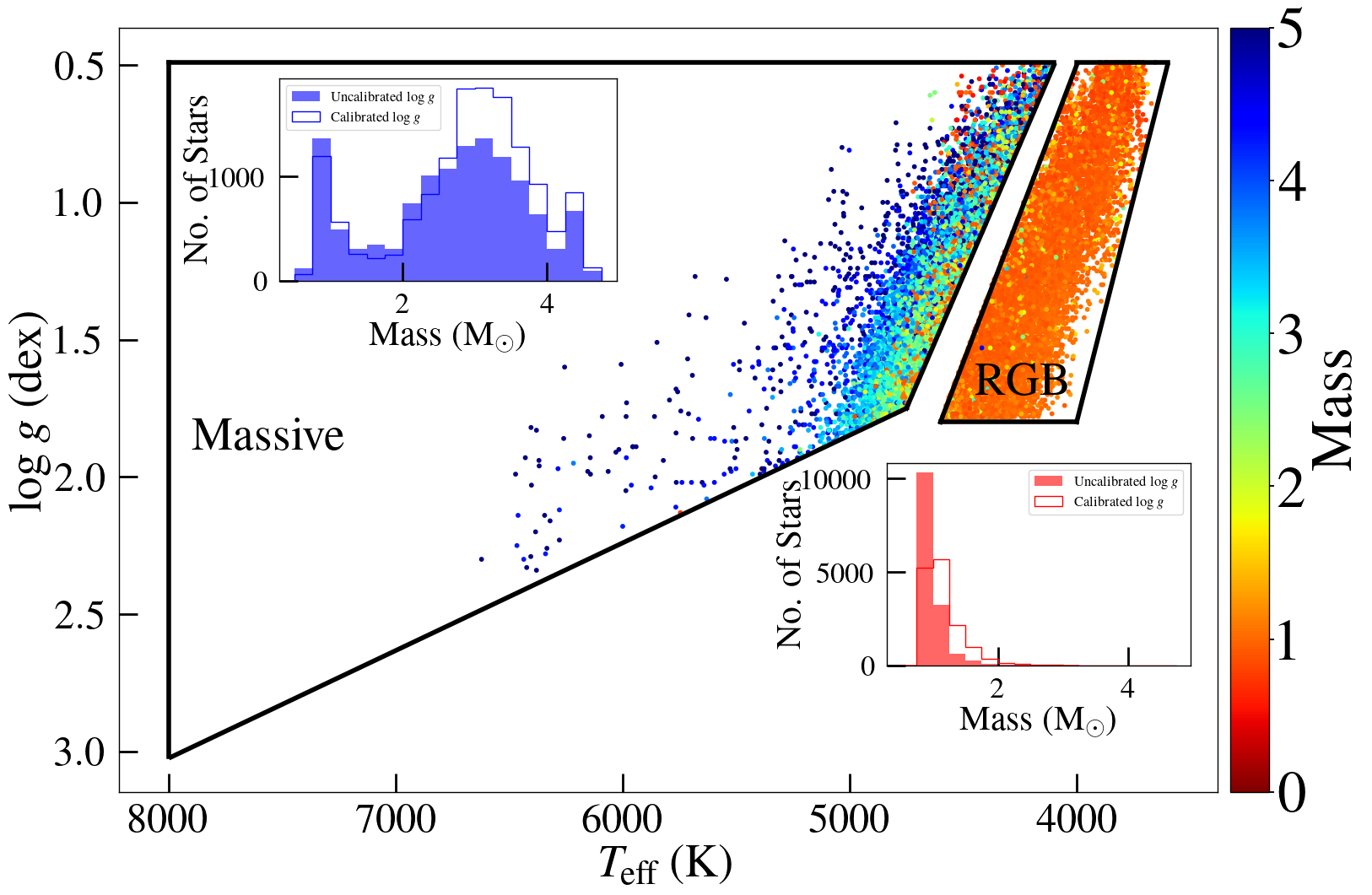}
  \caption{ 
  Kiel diagram (\teff\ versus \logg) based on uncalibrated Gaia DR3 spectroscopic parameters, showing the positions of 16\,881 massive sequence stars and 14\,799 RGB stars, colour-coded by their derived masses. The black boxes outline the approximate selection regions used to identify the two stellar populations used in \citet{Recio-Blanco:2023chemcart}. Masses are 
  the ones computed in this work, obtained combining spectroscopic and IR photometric data (see Sect.~\ref{sec:masses}).
  The inset panels display the mass distributions for massive sequence stars (blue) and RGB stars (red); filled and open histograms correspond to masses based on uncalibrated and calibrated \logg, respectively.
  }
  \label{fig:kiel_massive_rgb}%
\end{figure*}

\section{Data sets}
\label{sec:dataset}

The primary data set used in this work is the \textit{Gaia} GSP-spec catalog released with  \textit{Gaia} DR3. In addition, we have used the APOGEE DR17 \citep{DR17}, the GALAH DR4 \citep{Buder:2025}, and the \textit{Gaia} CNN \citep{Guiglion:2024} data sets to carry out detailed comparison of stellar parameters and selected elemental abundances of common stars in the \textit{Gaia} GSP-Spec sample. Below, we provide brief summaries of the data sets used in this work. 

\subsection{Gaia DR3}
\label{sec:gaiagspspec}
\vspace{-5pt}
Since massive sequence stars in the \textit{Gaia} RVS are the focus of this work, our main data set is the \textit{Gaia} GSP-Spec sample in the \textit{Gaia} DR3 version which provides the astrophysical parameters for 5.6 million stars. As mentioned in the introduction, the Gaia RVS spectra, covering a wavelength range of 845--872 nm and spectral resolution of 11,500, is analysed using the GSP-Spec module to derive the stellar parameters and elemental abundances of up to 11 elements. The details of the GSP-Spec pipeline are described in \cite{Recio-Blanco:2023} and the data is taken from the \textit{astrophysical$\_$parameters} table of \textit{Gaia} DR3. In this work, we use the global metallicities, mh$\_$gspspec, provided by \textit{Gaia} and we denote it as \feh\, in the text and the figures. To account for the biases in the raw GSP-spec \logg\, and \feh\, when compared to the literature, \cite{Recio-Blanco:2023} recommended raw \logg-based calibrations for \logg\, and \feh\, in their Table\,3. Meanwhile, both \logg-based and \teff-based calibrations for calcium abundances, [Ca/Fe], have been recommended in their Table\,4 assuming that the solar neighbourhood stars with metallicities close to zero and velocities close to the Local Standard of Rest should exhibit solar abundance distribution. In \cite{Recio-Blanco:2024}, new \teff-based calibrations have been recommended for \logg\, and metallicities in their Table\,A.1. We note that in this work we applied the recommended \teff-based calibrations for \logg\,, \feh, and [Ca/Fe] even though both \cite{Recio-Blanco:2023chemcart} and \cite{Spitoni:2023} use \logg-based calibrations.

\subsection{APOGEE}
\label{sec:apogee}

APOGEE is a near-infrared spectroscopic survey of the Milky Way observing in the H-band (15000-17000 \AA) at spectral resolution of R$\sim$22,500. The APOGEE-2 survey, as a part of SDSS-IV \citep{Blanton:2017}, covers both the northern and southern hemispheres with APOGEE instrument \citep{Wilson:2012} on the Sloan 2.5\,m Telescope \citep{Gunn:2006} at the Apache Point Observatory (APO), and the 2.5\,m du Pont telescope at Las Campanas Observatory \citep[LCO;][]{Bowen:1973} with the twin near-infrared (NIR) spectrograph \citep{Wilson:2019}, respectively. We use the latest APOGEE data release \citep[DR17;][]{DR17} that provides stellar parameters and elemental abundances of 20 species for 733,901 stars determined with the APOGEE Stellar Parameters and Chemical Abundances Pipeline \citep[ASPCAP;][]{ASPCAP:2016}, using MARCS stellar atmosphere model grids \citep{marcs:08}. In DR17, new synthetic spectral grids were created taking advantage of the new, non-local thermodynamic equilibrium (NLTE) population calculations by \cite{Osorio:2020} for Na, Mg, K, and Ca. In this work, we use the global metallicities, \feh, provided by APOGEE. 
APOGEE provides calibrated \teff\, and \logg\, to account for the deviation of the raw spectroscopic \teff and \logg\, from those determined by independent methods (photometric \teff\, and astroseismic \logg). Calibrated abundances have also been provided by addition of a zero point in order to bring the median abundance of the solar neighborhood solar metallicity stars to zero. In this work, we use the calibrated APOGEE stellar parameters and abundances.

\subsection{GALAH}
\label{sec:galah}
\vspace{-5pt}
GALAH is an optical spectroscopic survey of the Milky Way that observes using the
High Efficiency and Resolution Multi-Element Spectrograph \citep[HERMES;][]{Barden:2010,Sheinis:2015} on the Anglo-Australian Telescope. GALAH observes from 4700\,\AA~ to 7900\,\AA~ in four wavelength bands with spectral resolution, R$\sim$28,000. We used the data from the latest data release, GALAH DR4 \citep{GalahDR4}, that provides the stellar parameters and chemical abundances of up to 32 species for 917,588 stars determined using neural networks trained on synthetic spectra grids computed with Spectroscopy Made Easy \citep[SME;][]{sme,sme_code}. To convert the iron abundance based metallicity, [Fe/H], provided in the GALAH DR4 catalog to global metallicity, \feh, we used the correlation of \cite{salaris:2006}, [M/H] = [Fe/H] + log(10$^{[\alpha/Fe]}$ x 0.694 + 0.306), where [$\alpha$/Fe] is taken to be the mean of abundances of oxygen, magnesium, silicon, calcium and titanium abundances.

\subsection{\textit{Gaia} CNN}
\label{sec:gaiacnn}
\vspace{-5pt}
\textit{Gaia} CNN refers to the catalog made by \cite{Guiglion:2024} that provides homogeneous atmospheric parameters and abundances for 886 080 \textit{Gaia} RVS stars. The atmospheric parameters and chemical abundances, namely metallicity and global  $\alpha$-abundances, have been derived using a hybrid convolutional neural network (CNN) that combines the Gaia DR3 RVS spectra, photometry, parallaxes, and XP coefficients. The CNN was trained with a high-quality training sample based on APOGEE DR17 labels. 

\section{Selection of massive and RGB stars}
\label{sec:selection}

In this section, we describe the selection of massive and RGB stars from the data sets summarised in Sect.~\ref{sec:dataset}.  First, we applied the criteria shown in  Figure~8 of \cite{Recio-Blanco:2023chemcart} to select, using ADQL queries listed in the same paper, the massive and RGB stars. We further trimmed down these samples by selecting only the stars with valid GSP-Spec calcium abundances. The resulting subsample was then crossmatched with the APOGEE, GALAH, and \textit{Gaia} CNN datasets. The number of stars in the massive and RGB samples corresponding to each catalog cross matched with the \textit{Gaia} DR3 samples is listed in the Table~\ref{table:statistics}. Since \textit{Gaia} CNN is trained on APOGEE labels and do not provide individual alpha element abundances, in the following sections we focus only on APOGEE and GALAH, while the comparison with \textit{Gaia} CNN is discussed in the Appendix~\ref{sec:gaiacnn}.


\begin{table}
\caption{Number of stars in the massive and RGB stars catalogs selected from the primary catalog, \textit{Gaia} DR3, and further cross-match with APOGEE, GALAH, and \textit{Gaia} CNN catalogs.}\label{table:statistics}
\begin{tabular}{| c | c | c | c | c |}
\hline
 Catalog & \textit{Gaia} & APOGEE  & GALAH  & \textit{Gaia} CNN    \\
 \hline
Massive  & 16,881 & 276 & 306 & 573 \\
RGB & 14,787 & 14,787 & 1304 & 8561 \\
\hline
\end{tabular}
\end{table}


\subsection{Gaia DR3}
\label{sec:gaiagspspec_selection}

We started off by selecting the medium quality sample from the \textit{Gaia} archive using the ADQL query (listing 2) reported in the Appendix B of \cite{Recio-Blanco:2023chemcart}. From this sample, we selected the massive and RGB \textit{Gaia} GSPspec sources based on their position in the Kiel diagram \citep[see Fig. 8 of][]{Recio-Blanco:2023chemcart}. 

To select massive sequence stars, we imposed the constraint \logg\, < 1.75 + (\teff\, - 4750)/520 to select stars hotter than the RGB, and \logg\, < (\teff\,- 4750)/2550 + 1.75 to restrict the sample to the giant regime.  We also excluded all sources hotter than 8,000\,K or with surface gravity \logg\, lower than 0.5. The \teff\, and \logg\, values are taken from the teff$\_$gspspec field and the logg$\_$gspspec field respectively without any calibration applied as done in \cite{Recio-Blanco:2023chemcart}. This resulted in the selection of 49,225 massive sequence stars. On cross-match with 2MASS and WISE catalogs with a search radius of 2 arcseconds in RA,DEC using \textsc{topcat} \citep{topcat}, we have 48,870 stars. For these we estimated masses based on isochrone projections using \textit{Gaia} stellar parameters and 2MASS, WISE photometric information in Section~\ref{sec:masses}. Since the low abundances for massive sequence stars were reported for calcium in \cite{Recio-Blanco:2023chemcart}, we cross matched this set of stars with the calcium abundance sample selected with the ADQL query (listing 5) reported in the Appendix B of \cite{Recio-Blanco:2023chemcart}. This resulted in 16,881 massive sequence stars with valid calcium abundances.

To select RGB stars, we applied the constraints \logg\, < (1.3 x \teff\,-4480)/400, \logg\, > (1.3 x \teff\,-4900)/600, and 0.5 < \logg\, < 1.8 approximated from the Fig. 8 of \cite{Recio-Blanco:2023chemcart}. After cross match with the calcium abundance sample based on \textit{Gaia} source id, we have more than 500,000 RGB stars. We further cross matched this sample with APOGEE DR17 catalog based on \textit{Gaia} source id to reduce computational time when determining masses (see section~\ref{sec:masses}). We selected only those APOGEE stars  for which the 0 (TEFF$\_$WARN), 1 (LOGG$\_$WARN), 2 (VMICRO$\_$WARN), 3 (M$\_$H$\_$WARN), 4 (ALPHA$\_$M$\_$WARN), 7 (STAR$\_$WARN), 16 (TEFF$\_$BAD), 17 (LOGG$\_$BAD), 18 (VMICRO$\_$BAD), 19 (M$\_$H$\_$BAD), 20 (ALPHA$\_$M$\_$BAD), 23 (STAR$\_$BAD) bits of the ASPCAPFLAG have not been set and signal-to-noise ratio (S/N) > 40, ending up with 14,799 RGB stars. The cross-match with 2MASS and WISE catalogs using \textsc{topcat} resulted in a final sample of 14,787 RGB stars. 

In Figure~\ref{fig:kiel_massive_rgb}, we show the Kiel diagram colour coded with the masses estimated in Section~\ref{sec:masses} for the selected massive sequence stars and RGB stars from the \textit{Gaia} DR3 catalog. The criteria used for the selection are also shown with black solid lines. The inset panels show the mass distributions for massive sequence stars (blue) and RGB stars (red) with the filled and open histograms representing the masses computed with uncalibrated and calibrated \logg\, values respectively. Among the massive sequence stars, the fraction of stars with masses less than 2 M$\odot$\footnote{assuming 2 M$\odot$ to be the rough upper limit for low mass stars } ($\frac{N_{M<2M\odot}}{N_{total}}$) is estimated to be $\sim$ 16 $\%$ when using calibrated \logg\, and 22 $\%$ when using uncalibrated \logg.

\subsection{APOGEE}
\label{sec:apogee_selection}
\vspace{-5pt}

To select massive sequence stars in APOGEE with reliable calcium abundances from the \textit{Gaia} DR3 catalog, we cross-matched the APOGEE DR17 catalog with the 16,881 massive sequence stars selected from \textit{Gaia} DR3 based on \textit{Gaia} source id, yielding 276 stars in common. For RGB stars, we adopted the same sample of 14,787 \textit{Gaia} DR3 RGB stars previously cross-matched with APOGEE (Section~\ref{sec:gaiagspspec}). In both samples, APOGEE stars were filtered using the ASPCAPFLAG criteria described in Section~\ref{sec:gaiagspspec_selection}.

\subsection{GALAH}
\label{sec:galah_selection}
\vspace{-5pt}

The cross-match of GALAH DR4 catalog with the 16,881 massive sequence stars with valid calcium abundances from \textit{Gaia} DR3 catalog based on \textit{Gaia} source id resulted in 306 stars. 
The cross-match with the 14,787 RGB stars resulted in 1,304 stars. 
In both the cases, we choose only GALAH stars with the GALAH stellar parameter flag, \textit{flag$\_$sp}, and GALAH calcium abundance flag, \textit{flag$\_$ca$\_$fe}, set to 0.

\section{Mass computation}
\label{sec:masses}

\subsection{Description of the code}
\label{sec:mass_code}

Masses were computed using an updated version of the \citet{Kordopatis:2023a} pipeline. The code takes into consideration an isochrone library, on which any combination of observables can be projected onto (e.g. \teff, \logg, metallicity, magnitudes). 

For this study, we used the PARSEC v.1.2s isochrones\footnote{Two additional isochrone libraries were also tested: BaSTI and MIST, without changing the conclusions of this paper.} \citep{Bressan2012}, as downloaded and described in Section 2.1 of \citet{Kordopatis:2023a}. on which we projected the spectroscopic \teff, \logg, metallicity and the de-redenned infrared absolute magnitudes  derived from 2MASS (J, H, Ks) and WISE (W1, W2). We did not consider Gaia magnitudes, as they are more sensitive to reddening uncertainties \citep[see also][]{Kordopatis:2025}.

Absolute magnitudes were computed using the standard distance modulus formula adopting the \citet{Bailer-Jones:2021} geometric distances 
and correcting for the extinction. The latter was obtained using the {\tt dustmaps} python tool \citep{Green:2018} and querying the \citet{Schlegel:1998} maps.  For each star, the adopted reddening, E(B-V), was the \citet{Schlegel:1998} one, corrected by its distance, as in \citet{Kordopatis:2015}. Finally, the transformation from E(B-V) to the extinction in a given band ($A_\lambda/A_0$), 
was done using the transformation coefficients provided from the PARSEC website\footnote{\url{https://stev.oapd.inaf.it/cgi-bin/cmd}}.
The associated uncertainties on $A_\lambda/A_0$ were obtained by computing the standard deviation of 100 Monte-Carlo realisations of E(B-V), where the distances were drawn according to the posterior distribution published by \citet{Bailer-Jones:2021}. The adopted uncertainty on the absolute magnitudes was then obtained as the quadratic sum of the uncertainty on the distance modulus and the uncertainty on $A_\lambda/A_0$. For the massive sequence stars sample, the median uncertainty for the absolute magnitudes in the J, H, $K_s$, W1, W2 bands is 0.02\,mag (regardless of the band), whereas the 95th percentile of the uncertainty distributions are 0.037, 0.059, 0.034, 0.164, 0.059\,mag, respectively.

\begin{figure}
  \includegraphics[width=0.49\columnwidth]{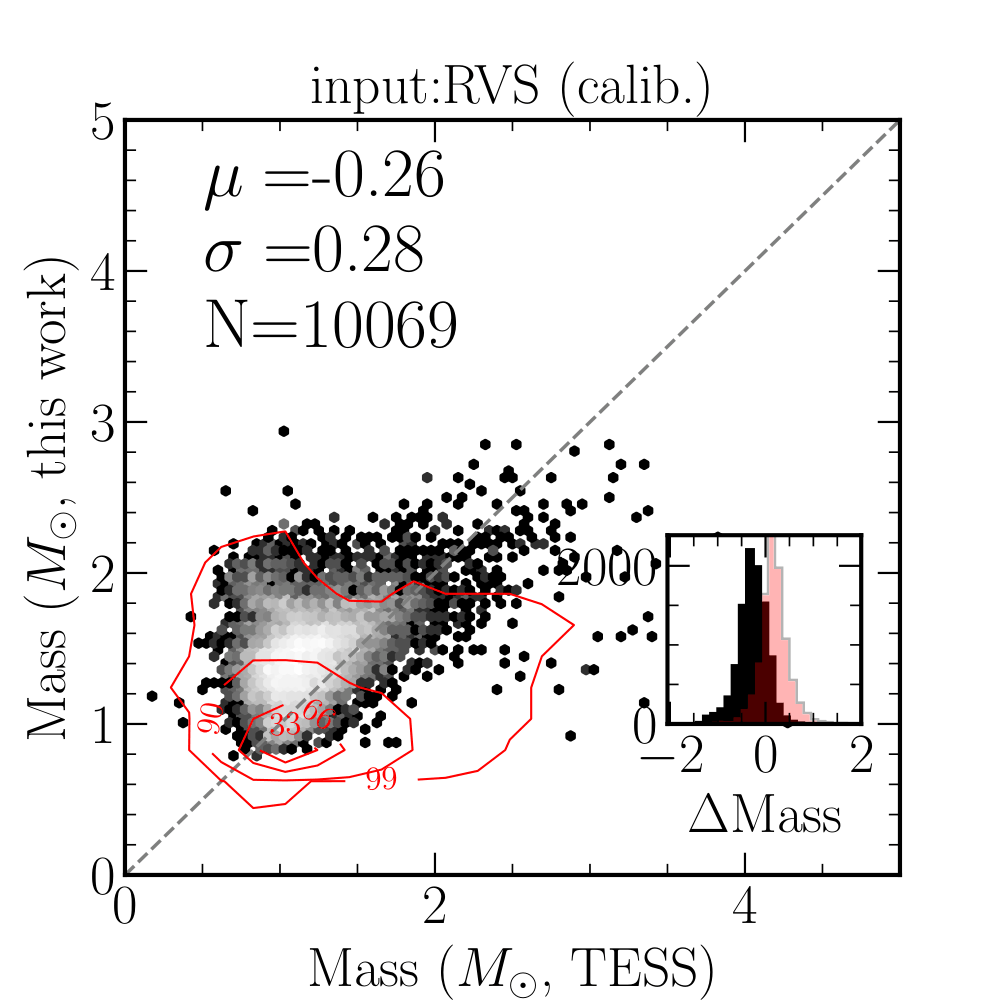}
    \includegraphics[width=0.49\columnwidth]{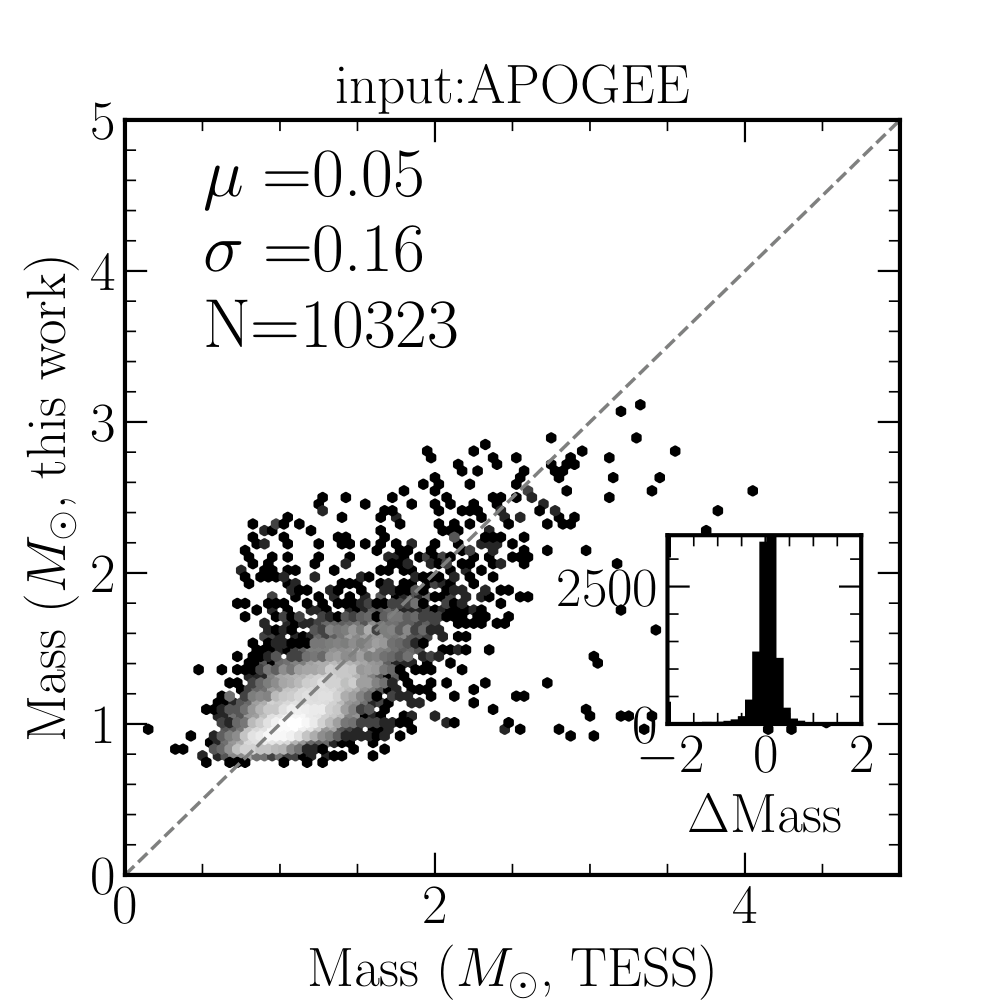} \\
     \includegraphics[width=0.49\columnwidth]{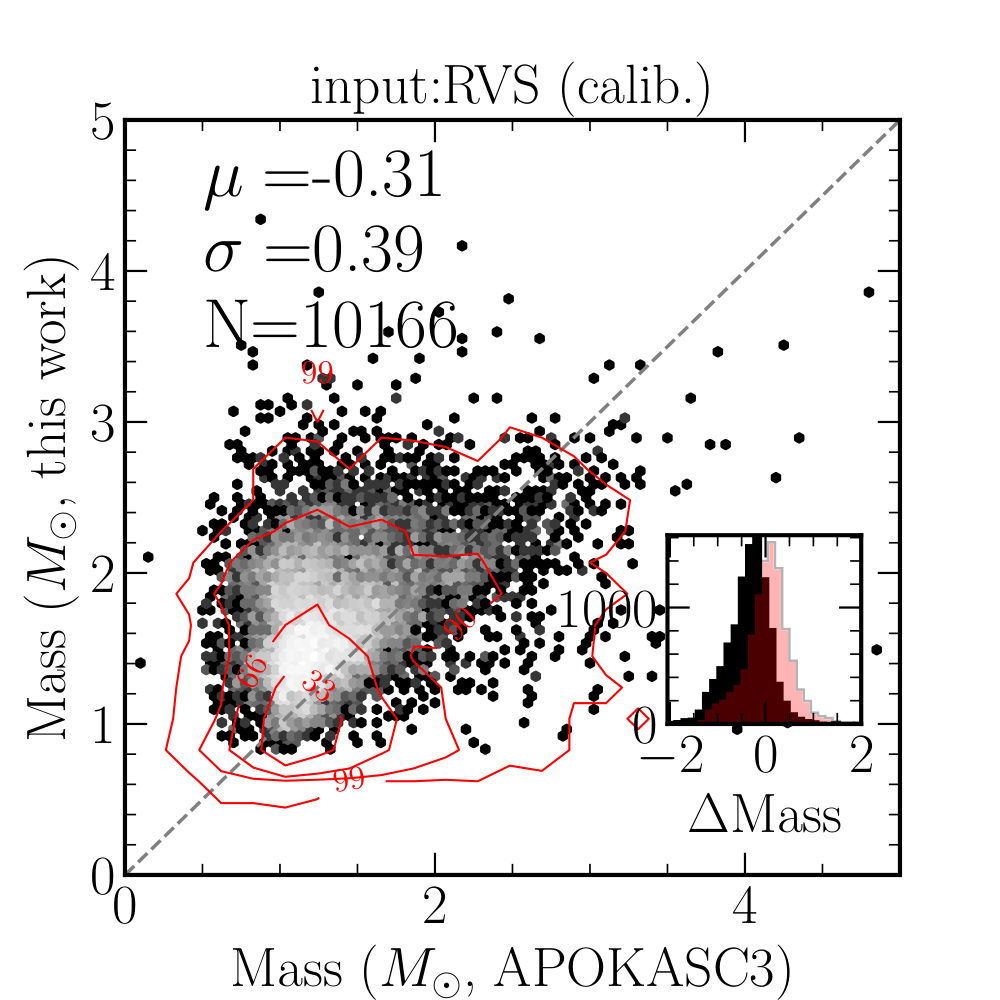}
    \includegraphics[width=0.49\columnwidth]{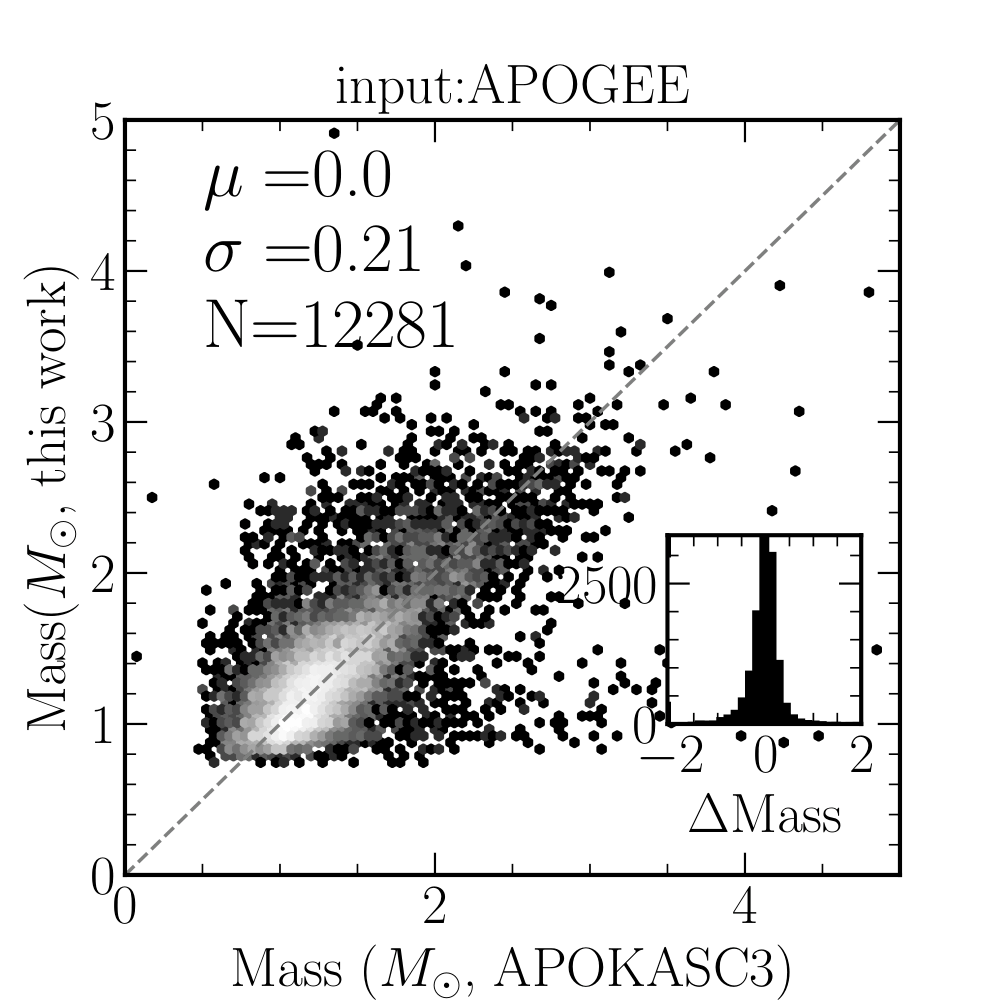}
  \caption{
  Comparison between asteroseismic masses (top: TESS; bottom: APOKASC-3) and the masses derived in this work using photometric and spectroscopic inputs projected onto PARSEC isochrones (see Sect.~\ref{sec:validation}). Left panels use GSP-Spec parameters (with red contours and inset distribution showing mass differences based on uncalibrated \logg), while right panels use APOGEE inputs. The median ($\mu$) and robust scatter ($\sigma$, scaled MAD) of the mass differences are indicated in each panel.
  }
  \label{fig:tess_APOKASC3}
\end{figure}

 \begin{figure*}
 \centering
  \includegraphics[width=0.9\textwidth]{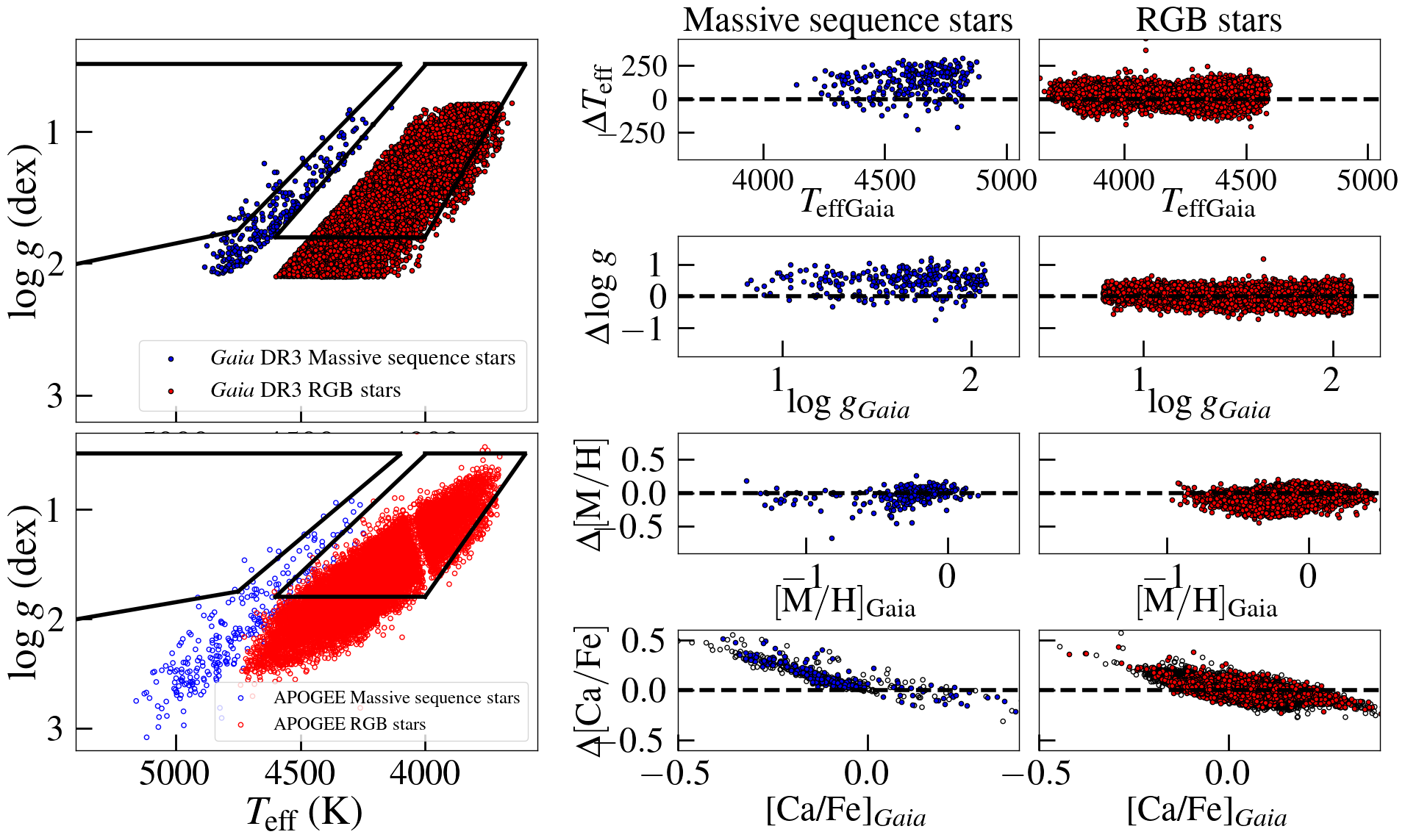}
  \caption{
  Comparison of stellar parameters and masses for massive (blue) and RGB (red) stars common to APOGEE and \textit{Gaia} DR3. Left panels show the Kiel diagrams with APOGEE (open circles) and \textit{Gaia} DR3 (filled circles); black boxes indicate the selection criteria adopted from \citet{Recio-Blanco:2023chemcart} (see Fig.~\ref{fig:kiel_massive_rgb}). Compared to the Fig.~\ref{fig:kiel_massive_rgb} stars lie outside the boxes in the case of \textit{Gaia} (upper panel) since we use calibrated \logg\, here. Middle and right panels show differences in stellar parameters and calcium abundances (APOGEE–\textit{Gaia}) as a function of the corresponding \textit{Gaia} DR3 values for massive and RGB stars, respectively. The differences in calcium abundances based on applying \logg\, calibrations in \textit{Gaia} are shown with black open circles. 
  }
  \label{fig:apg_gaia_diff}%
\end{figure*}

 \begin{figure*}
 \centering
  \includegraphics[width=0.9\textwidth]{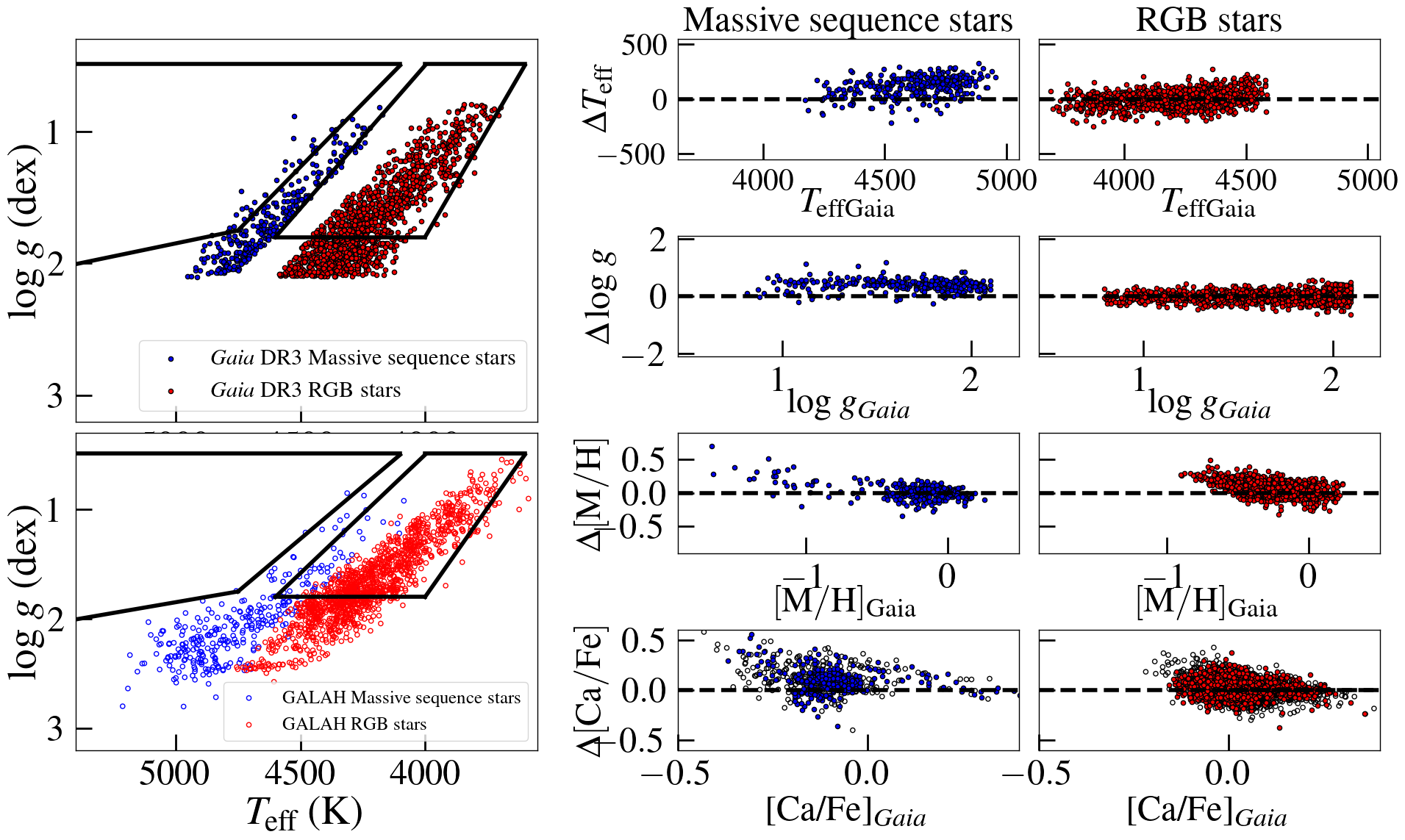}
  \caption{Similar to the  Figure~\ref{fig:apg_gaia_diff}, but showing the comparison of the stellar parameters and calcium abundances between the massive (blue) and RGB (red) stars in common between GALAH and \textit{Gaia} DR3.  }
  \label{fig:galah_gaia_diff}%
\end{figure*}

\subsection{Validation of the masses on APOKASC3 and TESS}
\label{sec:validation}
In order to validate the mass estimates of the selected stars, we compare the output of the code with  masses derived by asteroseismology \citep{Brown:1991,Kjeldsen:1995}. To do this, we choose the latest version of the APOKASC catalog, APOKASC-3 \citep{Pinsonneault:2025}, that contains the evolutionary state, asteroseismic surface gravity, mass, radius, age (and the APOGEE and Kepler data used to derive them) for 12,418 stars. After cross match with the APOGEE DR 17 catalog (based on 2MASS id), WISE catalog (with a search radius of 2 arcseconds using \textsc{topcat}, and the catalog of \textit{Gaia} DR3 astrophysical parameters (based on \textit{Gaia} source$\_$id), we have 12,364 stars with valid parameters to compute masses in the APOKASC3 catalog.  
As another comparison sample, we choose the catalog of masses and ages estimated for  132,794 giants \citep{Theodoridis:2025} selected from the NASA’s Transiting Exoplanet Survey Satellite \citep[TESS;][]{Ricker:2010} catalog of confirmed identified oscillators by \cite{Hon:2021}. among which 10,363 have also APOGEE atmospheric parameters. To obtain the final TESS sample for which masses can be estimated, we first select giants with signal-to-noise ratio higher than 150 in the APOGEE DR17 catalog leading to 1 80 421 stars. The APOGEE sample is cross matched with the version 8.2 of the TESS input catalog using \textsc{topcat} with a search radius of 2 arcseconds resulting in 1 80 407 stars. The stars in the TESS catalog is then cross matched with the above mentioned sample based on their TIC id leading to 10 363 stars that are also in the catalog of \textit{Gaia} DR3 astrophysical parameters (cross matched using their \textit{Gaia} source$\_$id).  

The photometric inputs (J, H, Ks, W1, W2) were combined with either the APOGEE or the GSP-spec atmospheric parameters, respectively (adopting either the calibrated or uncalibrated \logg\, for the latter), and projected onto the isochrones.

The results, shown in Fig.\,\ref{fig:tess_APOKASC3}, suggest that when  using the APOGEE inputs (right plots) our code manages to retrieve properly the stellar masses. The median difference compared to the asteroseismic measurements is smaller than $+0.05\,M_\odot$, whereas the robust standard deviation of the difference is of the order of 0.19$M_\odot$. Furthermore, the code converged for 10,323 stars and 12,281 stars of the TESS and the APOKASC3 samples, respectively (out of the 10,363 for TESS and 12,364 for the APOKASC3 catalogues). 

Using the GSP-spec parameters, the robust standard deviation of the difference is roughly twice larger and the bias is of the order of $-0.29\,M_\odot$ (i.e. masses are over-estimated), due to the simultaneous effect of less accurate and less precise parameters (which is attributed, in turn,  to the lower resolution and smaller wavelength range of the RVS, see for example \citealt{Kordopatis23b}). One can also notice that the projections obtained with GSP-spec parameters converged for less stars than with the APOGEE parameters (10,069 stars for TESS and 10,166 stars for APOKASC3), indicating that less isochrones have been found to match the given inputs.  

Finally, we note that the projection using GSP-spec parameters presents a high-mass plume\footnote{Using the calibrations based on \logg, while removing the overall bias, do not decrease the scatter, nor remove the high-mass plume. } for stars that have asteroseismically derived masses of $\sim 1 M_\odot$. Hints of the latter is also observed - but to a much lesser extent - for the APOGEE projection.


\section{Comparison between \textit{Gaia} DR3 and other catalogs}
\label{sec:comparison}

In this section, we investigate the differences in stellar parameters (\teff, \logg, \feh) and calcium abundances between \textit{Gaia} DR3 and APOGEE, GALAH for the common stars in the massive and RGB star samples listed in the Table~\ref{table:statistics}. We also compare the [Ca/Fe] versus [M/H] plots as well as the mass histograms with the masses computed using the respective stellar parameters (\teff, \logg, and \feh) from \textit{Gaia} DR3, APOGEE and GALAH. We do similar investigations using the cross-matched samples from the \textit{Gaia} CNN catalog in Appendix~\ref{sec:gaiacnn}. Further, we investigate whether the cross-matched subsets of low calcium stars in APOGEE and GALAH are representative (in terms of magnitude ranges, spatial coverages, stellar parameters etc) of the \textit{Gaia} parent sample of low calcium massive sequence stars in the Appendix~\ref{sec:representativeness}. 

\subsection{Differences in stellar parameters and calcium abundances}
\label{sec:comparison_stellarparameters}

\begin{figure*}
 \centering
  \includegraphics[width=0.9\textwidth]{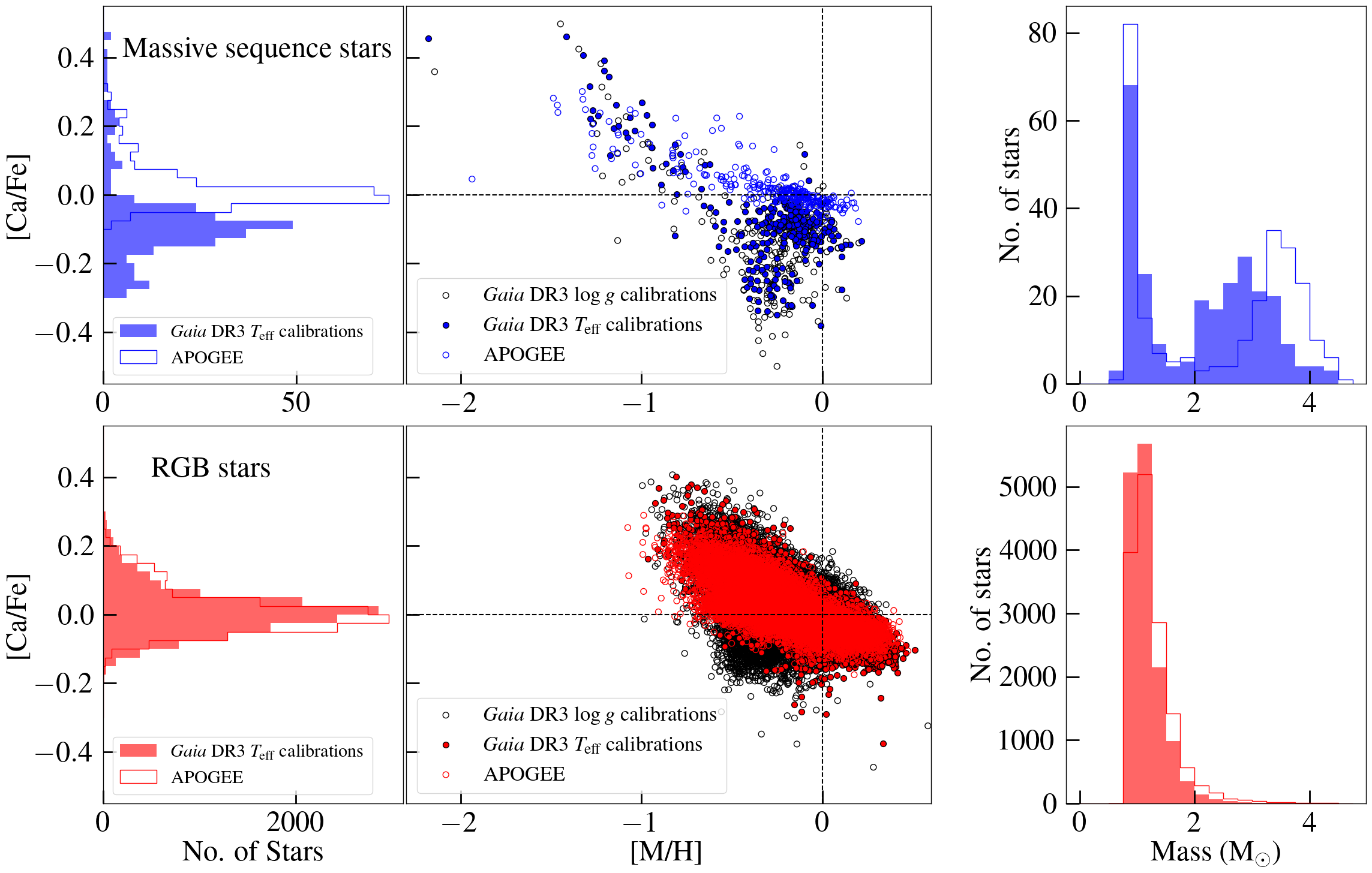}
  \caption{
  [Ca/Fe] vs M/H] trends for massive (blue, top middle panel) and RGB (red, bottom middle panel) sources in common between APOGEE (open circles) and \textit{Gaia} DR3 (filled circles). The filled circles and black open circles represent \textit{Gaia} sources based on \teff\, calibration and \logg\, calibration respectively. Left panels show the [Ca/Fe] distributions from APOGEE (open histograms) and \textit{Gaia} DR3 (filled histograms) for massive (top) and RGB (bottom) stars. Right panels show the corresponding mass distributions, with masses computed using \textit{Gaia} GSP-spec parameters shown as filled bars and those computed using each survey’s own stellar parameters shown as open bars.
  }
  \label{fig:apg_gaia_validca_diff}%
\end{figure*}

\begin{figure*}
 \centering
  \includegraphics[width=0.9\textwidth]{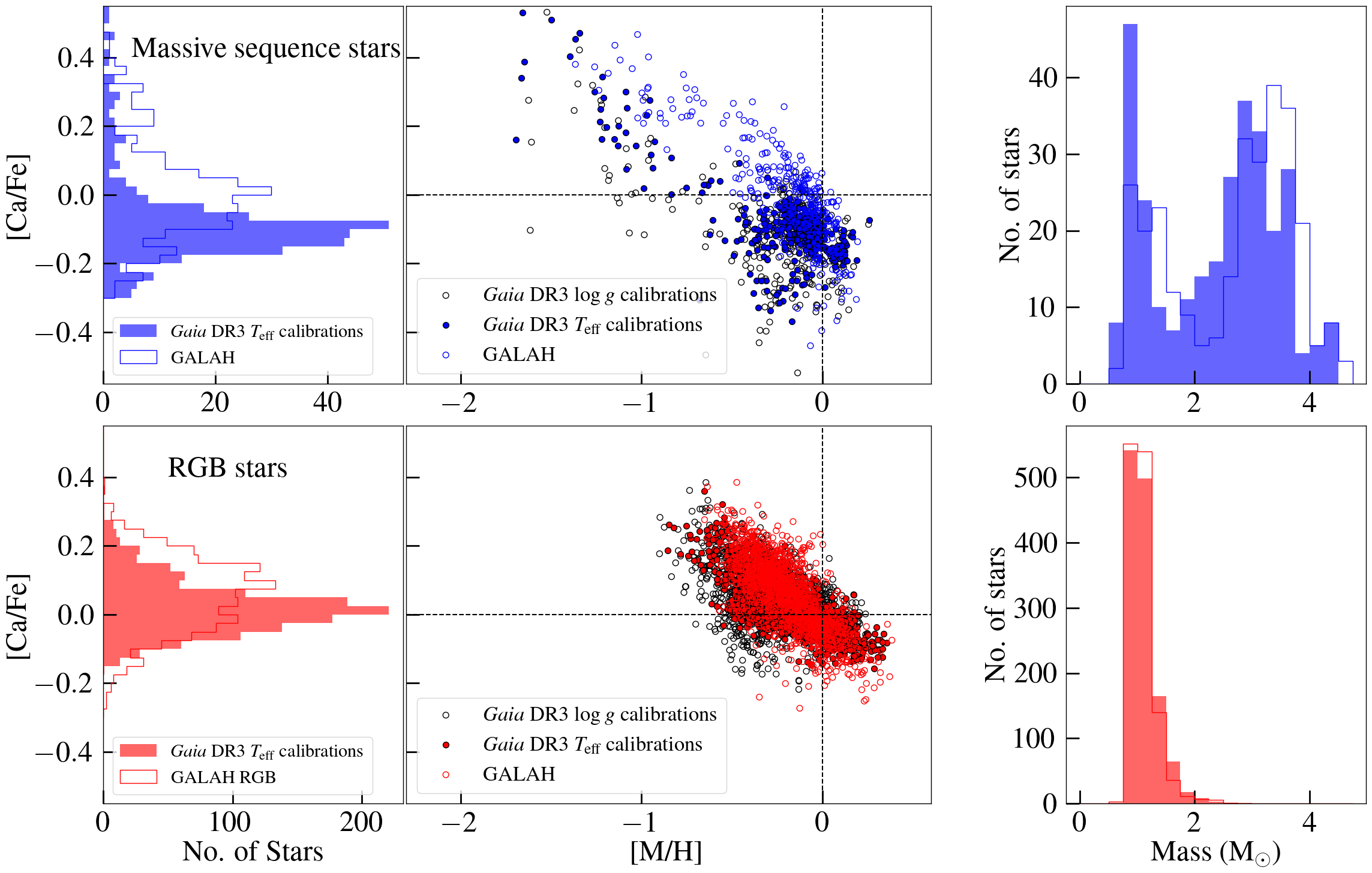}
  \caption{Similar to the Figure~\ref{fig:apg_gaia_validca_diff}, but for the common stars in GALAH and \textit{Gaia} DR3. }
  \label{fig:galah_gaia_validca_diff}%
\end{figure*}

In Figures~\ref{fig:apg_gaia_diff} and \ref{fig:galah_gaia_diff}, we show the Kiel diagrams (\logg\, versus \teff) for the massive (red) and RGB (blue) stars in common between \textit{Gaia} DR3 (closed circles) and the respective external catalogs (open circles) in the left panels. The plots in the middle and right columns show the differences in stellar parameters and calcium abundances for the common stars in the massive sequence stars sample and RGB sample between APOGEE and \textit{Gaia} (APOGEE-\textit{Gaia} on y-axis) and GALAH and \textit{Gaia} (GALAH-\textit{Gaia} on y-axis) respectively. We use the calibrated values of stellar parameters and calcium abundances in the case of APOGEE, and as mentioned in the Section~\ref{sec:gaiagspspec} we applied the \teff-based calibrations for \logg\,, \feh, and [Ca/Fe] in the case of \textit{Gaia} DR3. We note that the \teff-based calibrations lead to less scatter in the calcium abundance trends and distributions for the massive and RGB stars in comparison with the \logg-based calibrations (see panels with calcium in Figures~\ref{fig:apg_gaia_diff}, ~\ref{fig:galah_gaia_diff}, ~\ref{fig:apg_gaia_validca_diff}, ~\ref{fig:galah_gaia_validca_diff}, and table~\ref{table:parameterdiff}). The black boxes in the Kiel diagrams represent the selection criteria shown in Figure~\ref{fig:kiel_massive_rgb} for massive and RGB stars. 
The mean differences and dispersions in the stellar parameters and calcium abundances are listed in the Table~\ref{table:parameterdiff}.

From the Kiel diagrams, it is evident that the majority of the selected massive sequence stars in APOGEE and GALAH lie outside the box defined for massive sequence stars (see lower left panels in the figures~\ref{fig:apg_gaia_diff} and \ref{fig:galah_gaia_diff}) with higher \teff\, and \logg\, values. Similar increase in \logg\, is seen in the case of \textit{Gaia} DR3 when calibrations are applied to the \logg\, values (see upper left panels in the figures~\ref{fig:apg_gaia_diff} and \ref{fig:galah_gaia_diff}) resulting in the majority of the selected massive sequence stars to lie outside the box which is the same as the one shown in the Figure~\ref{fig:kiel_massive_rgb}. At the same time, the majority of the selected RGB stars in all the three external catalogs and \textit{Gaia} DR3 (with calibrated \logg) lie within the box defined for RGB stars, and the stars outside the box have generally higher \logg\, values. 

The \teff, and  \logg\, values of the massive sequence stars in \textit{Gaia} DR3 are different from those in APOGEE and GALAH (see the middle column panels of the figures~\ref{fig:apg_gaia_diff} and \ref{fig:galah_gaia_diff}). Specifically, the APOGEE and GALAH \teff\, values are higher with the mean difference in \teff\, above 100 K and dispersion close to 100 K. Similarly, the APOGEE and GALAH \logg\, values are higher than \textit{Gaia} DR3 calibrated values with a mean difference of $\sim$0.4-0.5 dex and significant dispersion of $\sim$0.2-0.3 dex. \feh\, values of massive sequence stars in the external catalogs are in good agreement with \textit{Gaia} DR3 values with mean differences of $\sim$ 0.02 dex and 0.13 dex dispersions.

Both external catalogs exhibit similarly higher [Ca/Fe] for massive sequence stars compared to \textit{Gaia} DR3 with mean differences of 0.09-0.12 dex and $\sim$ 0.12-0.13 dex dispersions. We also find, for the external catalogs, a downward trend in the [Ca/Fe] abundances versus the \textit{Gaia} abundances for majority of stars, such that the differences are larger for lower Gaia abundances. There is better agreement at higher values of \textit{Gaia} DR3 [Ca/Fe] ($\sim$ 0.0 dex) and gradually increasing differences (upto $\sim$ 0.6 dex) with decrease in \textit{Gaia}DR3 [Ca/Fe]. To summarise, the three external catalogs (including \textit{Gaia} CNN discussed in Appendix~\ref{sec:gaiacnn}) determine higher [Ca/Fe] values for the massive sequence stars (at high metallicities) for which \textit{Gaia} DR3 estimate lower [Ca/Fe] values.

For the RGB stars, the mean differences and the dispersions in stellar parameters and calcium abundances are significantly lower than those of massive sequence stars, except for \feh\, for which the differences are comparable (see the right most column panels of the figures~\ref{fig:apg_gaia_diff} and \ref{fig:galah_gaia_diff}). Specifically, the mean difference in \teff\, is higher for the external catalogs with mean differences lower than 40\,K and dispersions of the order of 45-70\,K. The \logg\, values are in much better agreement with mean differences smaller than -0.03 dex and dispersions of the order of 0.17\,dex. The mean differences and dispersions in \feh\, are in the range of -0.04 to 0.0\,dex and 0.07 to 0.11\,dex, respectively. The agreeement in [Ca/Fe] between the external catalogs and \textit{Gaia} DR3 is better for RGB stars with mean differences of 0.0 to 0.04 dex and dispersions of 0.04 to 0.08 dex. In APOGEE (and \textit{Gaia} CNN discussed in Appendix~\ref{sec:gaiacnn}), there is a hint of the downward trend in $\Delta$[Ca/Fe] with increase in \textit{Gaia} DR3 [Ca/Fe] seen in the case of massive sequence stars, but this is not seen in GALAH. This may be attributed to the larger scatter in calcium abundances in GALAH. Overall, there is better agreement in stellar parameters and calcium abundances between \textit{Gaia} DR3 and the external catalogs for the selected RGB stars.

 \begin{figure*}
 \centering
  \includegraphics[width=0.9\textwidth]{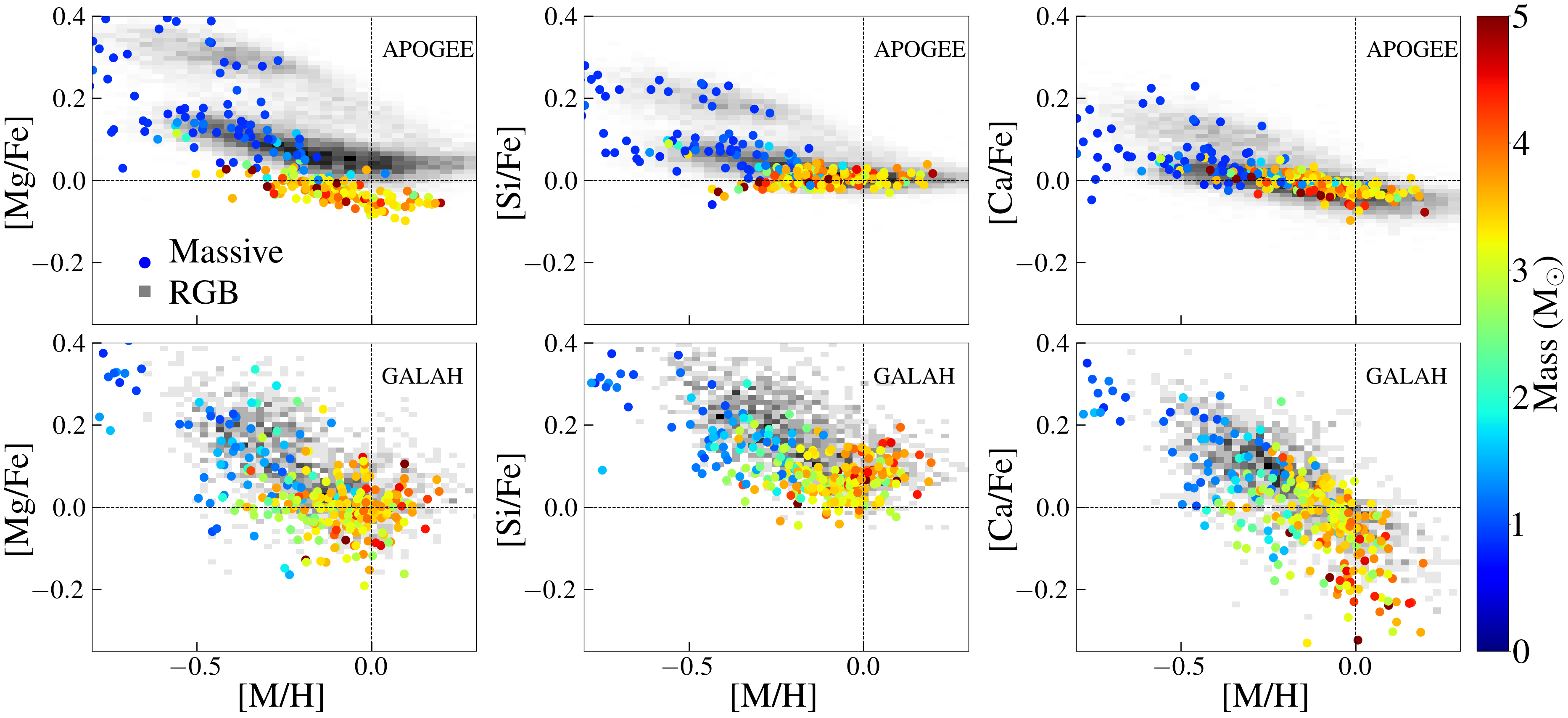}
  \caption{ Mg, Si, and Ca abundance trends versus metallicity for massive (filled circles) and RGB stars (gray 2D histograms) in APOGEE (top panels) and GALAH (bottom panels). The massive sequence stars are color-coded with the masses estimated in this work using the stellar parameters from APOGEE and GALAH as inputs. }
  \label{fig:apogee_galah_mgsica}
\end{figure*}

\begin{table}
\caption{ The mean and standard deviation of the differences in \teff, \logg, \feh\,, and [Ca/Fe] between the sources in common between \textit{Gaia} DR3 and the three external catalogs for the respective masssive and RGB star samples. The values in brackets are the mean and standard deviation in calcium abundances when using \logg\, based calibrations for \textit{Gaia}.}\label{table:parameterdiff}
\begin{tabular}{|c | c | c c c c |}
 \hline
Catalog &  & $\Delta$\teff & $\Delta$\logg & $\Delta$\feh & $\Delta$[Ca/Fe]    \\

\hline
\multirow{4}{*}{APOGEE}    & $\mu_{massive}$  & 134 & 0.46 & -0.02 & 0.12 (0.14)  \\
 & $\sigma_{massive}$  &  96 &  0.32 &  0.11 & 0.12 (0.14) \\
  & $\mu_{RGB}$ & 32 & 0.03 & -0.04 & 0.0 (-0.02) \\
 & $\sigma_{RGB}$ &  45 &  0.17 &  0.07 & 0.04 (0.06)  \\
\hline 
\multirow{4}{*}{GALAH}   & $\mu_{massive}$  & 118  & 0.39  &  0.02 & 0.09 (0.11) \\
& $\sigma_{massive}$ &  93  &  0.20  &   0.13 & 0.13 (0.13) \\
& $\mu_{RGB}$ & 3  & 0.0  &  0.0 & 0.04 (0.01) \\
& $\sigma_{RGB}$  &  67  &  0.17  &   0.11 &  0.08 (0.1)  \\
\hline 
\end{tabular}
\end{table}

\subsection{Calcium abundance trends and masses }
\label{sec:comparison_calciumtrend}

In the middle panels of Figures~\ref{fig:apg_gaia_validca_diff} and \ref{fig:galah_gaia_validca_diff} we show the [Ca/Fe] versus [M/H] trends for the massive (blue) and RGB (red) stars in common between \textit{Gaia} DR3 (closed circles) and the respective external catalog (open circles). The left panels of each figure show the [Ca/Fe] distributions of the massive (blue) and RGB (red) stars with the \textit{Gaia} DR3 [Ca/Fe] represented by filled bars and the respective survey [Ca/Fe] represented by open bars. The right panels in each figure show the mass distributions of the massive (blue) and RGB (red) stars with the masses computed using the \textit{Gaia} DR3 stellar parameters represented by the filled bars and those computed using the stellar parameters from the respective external catalog represented by the open bars. 

The [Ca/Fe] versus [M/H] trends and [Ca/Fe] distributions reiterate the differences between \textit{Gaia} DR3 and the external catalogs shown in the Figures~\ref{fig:apg_gaia_diff} and \ref{fig:galah_gaia_diff}. For the massive sequence stars, the [Ca/Fe] distributions of the external catalogs peak around 0.0 dex with low scatter in APOGEE and more scatter in GALAH. Meanwhile \textit{Gaia} DR3 [Ca/Fe] distributions of the common massive sequence stars peak at sub-solar [Ca/Fe] ($\sim$-0.1 to -0.2 dex) and exhibit larger scatter skewed towards lower [Ca/Fe] values. 
The mass distributions of the selected massive sequence stars in the \textit{Gaia} DR3 as well as the external catalogs have two peaks, one around 1 M$_{\odot}$ and the other around 3-4 M$_{\odot}$. We note that \textit{Gaia} DR3 masses for the higher mass peak sample has a slight offset to higher values compared to APOGEE and GALAH masses. The presence of lower mass ($<$ 2 M$_{\odot}$) stars among the selected massive sequence stars indicates contamination by low mass RGB stars due to the adopted massive sequence stars selection criteria (see Figures~\ref{fig:apg_gaia_diff} and \ref{fig:galah_gaia_diff}).

For RGB stars, the [Ca/Fe] versus [M/H] trends show similar scatter for the two external catalogs as well as \textit{Gaia} DR3 (see Appendix~\ref{sec:gaiacnn} for discussion on comparsion with \textit{Gaia} CNN) . The [Ca/Fe] distributions of \textit{Gaia} DR3 as well as the external catalogs peak close to 0.0 dex and above. Similarly, there is clear agreement between the masses estimated using the \textit{Gaia} DR3 stellar parameters and the respective external catalog stellar parameters. All the mass distributions are limited to less than 2 M$_{\odot}$ and peaks between 1 and 2 M$_{\odot}$. This indicates that the selection box for the RGB stars is able to filter low mass RGB stars with least contamination even though a significant portion of the stars lie outside the selection box (see the left panels in the Figures~\ref{fig:apg_gaia_diff} and \ref{fig:galah_gaia_diff}).

 \subsection{Abundances of other elements }
\label{sec:comparison_otheralpha}

Up to this point, our analysis has focused on calcium, motivated by the low calcium abundances reported for massive sequence stars by \citet{Recio-Blanco:2023chemcart}. Using three independent external catalogs for comparison, we have shown that their calcium abundances are consistent with those of low-mass RGB stars and do not indicate any depletion. However, as noted in the introduction, \citet{Spitoni:2023} demonstrated that a subset of APOGEE massive sequence stars (overlapping with the \textit{Gaia} DR3 massive-star sample) exhibits  lower magnesium abundances than low-mass RGB stars.

This raises the question of whether our selected samples of massive sequence stars from APOGEE and GALAH exhibit lower abundances in any other $\alpha$-element species, i.e., magnesium and silicon. To investigate this, we represent in Fig.~\ref{fig:apogee_galah_mgsica} the abundance trends as a function of metallicity for two additional $\alpha$-elements: magnesium (left panels) and silicon (middle panels). The comparison is shown for the massive sequence stars (filled circles) and the RGB stars (blue contours) in APOGEE (top row) and GALAH (bottom row). Massive sequence stars are color-coded by the masses estimated based on each survey’s stellar parameters in this work. We do not include abundances for other $\alpha$-elements from the \textit{Gaia} DR3 catalog, since our massive and RGB samples were selected primarily on the requirement of having reliable calcium abundances (see Section~\ref{sec:selection} for details).

In Fig.~\ref{fig:apogee_galah_mgsica}, the higher-mass stars ($> 2,\mathrm{M}_\odot$; yellow and red filled circles) in both the APOGEE and GALAH samples occupy metallicities above $\sim -0.5$ dex and fall within the low-$\alpha$ population for all three elemental abundance trends. In GALAH, these high-mass stars show magnesium, silicon, and calcium abundances consistent with those of the majority of RGB stars at similar metallicities. The same holds for APOGEE, with the exception of magnesium.

For APOGEE, the lower-mass stars ($< 2,\mathrm{M}_\odot$; blue filled circles) in the massive-star sample follow the [Mg/Fe] sequence traced by the RGB stars (blue contours) down to metallicities of approximately $-0.2$ dex. In contrast, the magnesium abundances of most high-mass APOGEE stars are sub-solar, showing a sharp drop to lower [Mg/Fe] compared to both the lower-mass APOGEE stars and the low-$\alpha$ APOGEE RGB population. As a result, the high-mass APOGEE stars form a distinct sequence of magnesium-poor stars relative to the low-mass RGB stars. This behavior is consistent with the findings of \citet{Spitoni:2023} (their Fig. 14). We investigate this feature in more detail in Section~\ref{sec:lowmg_apogee}.

\section{Discussion}
\label{sec:discussion}

In this section, we discuss the results from our investigation into the chemically depleted massive star population in the solar neighborhood identified by \cite{Recio-Blanco:2023chemcart}. 
We begin with a brief note on the contamination of the massive-star sample by low-mass stars arising from our selection criteria. We then assess the presence or absence of a chemically depleted massive population in the \textit{Gaia} DR3 catalog and in the three external catalogs, drawing on the detailed comparisons performed between stellar parameters, $\alpha$-element abundances, and masses for the massive and RGB samples across these datasets. Next, we examine the origin of the low magnesium abundances observed in a subset of the APOGEE massive-star sample. Finally, we discuss the implications of our results for the three-infall chemical evolution model proposed by \citet{Spitoni:2023} to explain the chemical depletion observed in young massive sequence stars in the solar neighborhood.

\subsection{Contamination in the selected samples } 
\label{sec:contamination}

As the first step in our investigation into the chemical impoversihment found in the \textit{Gaia} DR3 massive star population, we selected the massive and RGB stars based on the selection boxes defined in the unclaibrated \textit{Gaia} DR3 \teff\, versus \logg\, plot shown in the Figure~\ref{fig:kiel_massive_rgb} (adopted from the Figure 8 in \citealt{Recio-Blanco:2023chemcart}). The masses we estimated using both calibrated and uncalibrated \textit{Gaia} DR3 \logg\, values revealed a contamination of $\sim$16$\%$ and $\sim$23$\%$ respectively of low mass stars ($<$ 2 M$_{\odot}$) among the selected massive sequence stars population, and negligible contamination of massive sequence stars among the selected RGB stars (see the inset histograms in Figure~\ref{fig:kiel_massive_rgb}). This indicates that the selection boxes employed are not the most efficient way to choose massive sequence stars especially using the uncalibrated \logg. This is further supported by the fact that majority of the massive sequence stars lie outside the massive sequence stars selection box in the \textit{Gaia} DR3 \teff\, versus \logg\, plots with calibrated \logg\, shown in the top left panels in the Figures~\ref{fig:apg_gaia_diff}, \ref{fig:galah_gaia_diff}, and \ref{fig:cnn_gaia_diff}. Yet, it is encouraging that more than 80$\%$ of the selected stars are indeed massive when using calibrated \logg\, values.

\subsection{Calcium lines in the \textit{Gaia} spectra}
\label{sec:gaiaspectra}

\begin{figure}
  \includegraphics[width=\columnwidth]{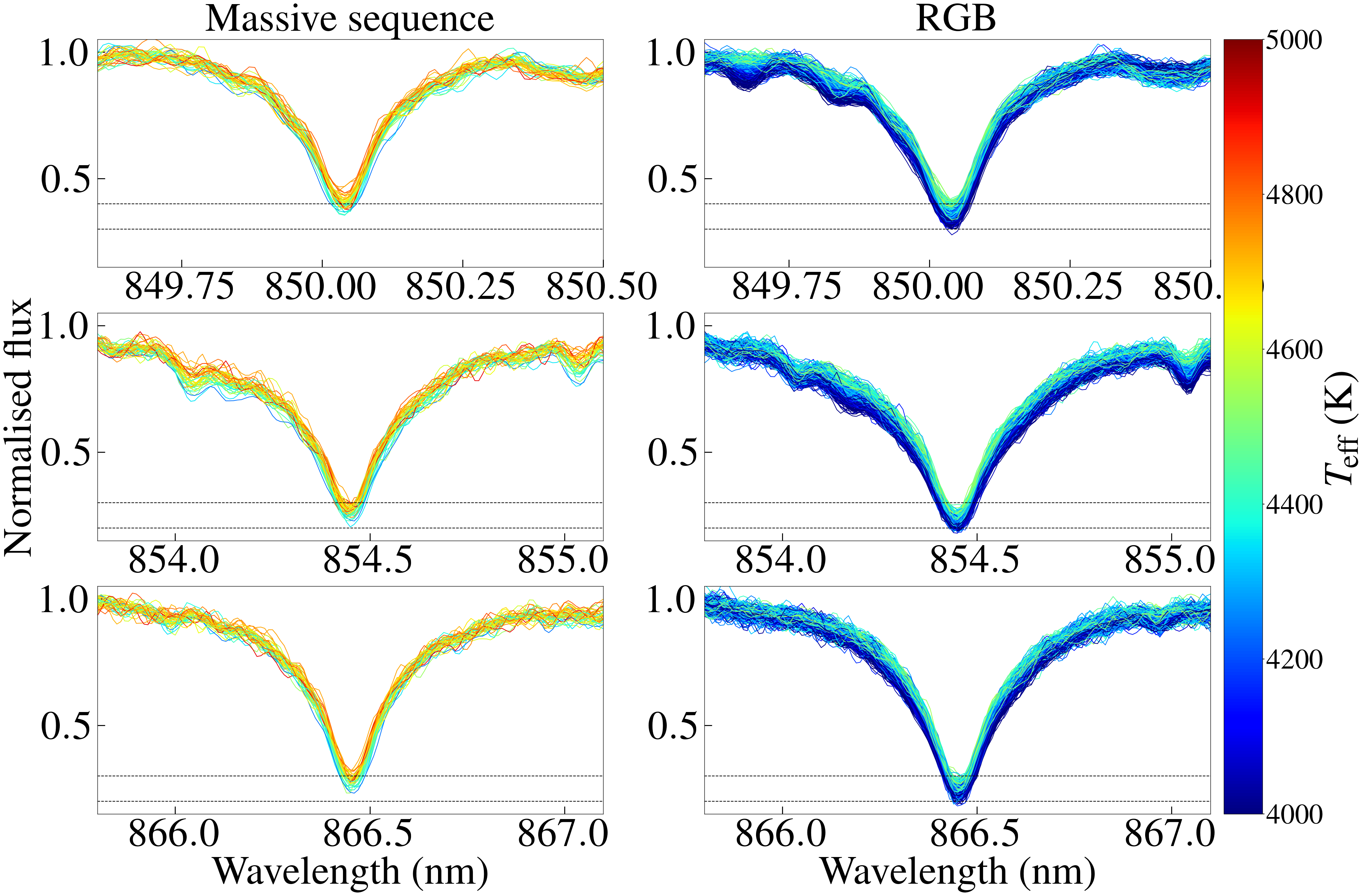}
  \caption{Spectral windows showing the caclium lines in the \textit{Gaia} spectra of low calcium massive sequence stars (-0.4 < [Ca/Fe] < -0.15) and RGB stars within the same metallicity range (-0.5 < \feh\, < 0.0 dex). The massive sequence stars spectra are plotted in the left column and the RGB stars spectra in the right column with each row corresponding to one calcium line. Horizontal dashed lines in each plot are there to guide the readers in visually sizing up the line depths.   }
  \label{fig:gaiaspectra_ca}
\end{figure}

In order to check for any obvious differences in the calcium lines in the spectra of the \textit{Gaia} massive sequence stars and RGB stars, we downloaded the spectra of few massive sequence stars and RGB stars. In particular, we selected massive sequence stars with lowest values of [Ca/Fe] (-0.4 < [Ca/Fe] < -0.15) and few randomly selected RGB stars. We also made sure that both samples lie within the same metallicity range (-0.5 < \feh\, < 0.0 dex) where the differences in calcium abundances between the two samples are the most prominent. In the Figure~\ref{fig:gaiaspectra_ca}, the spectral windows covering the three calcium lines are plotted side-by-side with massive sequence stars in the left column and RGB stars in the right column color coded with their \textit{Gaia} \teff. Horizontal dashed lines in each row are at the same levels and are plotted to guide the readers in visually sizing up the line depths. We note that the calcium line depths in the massive sequence stars are smaller than in the RGB stars. At the same time, the color coding with \teff\, clearly indicate that this is due to the \teff-dependence of line strengths, and thus the weaker calcium lines seen in the massive sequence stars spectra is a \teff\, effect.
Furthermore, although massive (and hence young) stars are known to exhibit stellar activity, we do not see any strong evidence of such in the spectra that could alter significantly the calcium abundance determination. We also checked the values of the stellar activity index, \textit{activityindex$\_$espcs}, determined from the analysis of the Ca II infrared triplet lines in the RVS spectrum \citep{Lanzafame:2023}, for the 16,881 stars in the \textit{Gaia} massive sequence stars sample. However, only nine stars have valid activity index measurements, which is insufficient to draw any meaningful conclusions.

\subsection{Is the massive sequence stars population chemically depleted? }
\label{sec:chemimpoverish}

To validate the \textit{Gaia} DR3 stellar parameters and calcium abundances for the massive and RGB samples, we cross-matched these stars with two Milky Way spectroscopic surveys with higher spectral resolution than \textit{Gaia} DR3: APOGEE and GALAH. We note that although the number of stars in the APOGEE and GALAH overlaps are small, they seem to cover the same regions of the CMD, Kiel diagram, and the sky as the parent \textit{Gaia} massive sequence sample, though the sky coverage present some limitations.  We also included the \textit{Gaia} CNN catalog, which predicts stellar parameters and overall $\alpha$-element abundances directly from the \textit{Gaia} spectra using CNN models trained on APOGEE labels. Including the \textit{Gaia} CNN catalog allows us to compare stellar parameters and abundances derived from the same underlying spectra as \textit{Gaia} DR3, but obtained using an entirely different methodology.

The comparisons shown in Figs.~\ref{fig:apg_gaia_diff}, ~\ref{fig:galah_gaia_diff}, and \ref{fig:cnn_gaia_diff} reveal significant offsets and large dispersions in $T_{\rm eff}$, $\log g$, and calcium abundances for massive sequence stars, with \textit{Gaia} DR3 values systematically lower than those from APOGEE and GALAH. In the case of the \textit{Gaia} CNN catalog, $T_{\rm eff}$ and $\log g$ show better overall agreement with \textit{Gaia} DR3 -though still with substantial scatter -while the $\alpha$-element abundances are higher than the \textit{Gaia} DR3 estimates, consistent with the trends seen in APOGEE.

In contrast with the massive sequence stars, the RGB stellar parameters and calcium abundances agree much better with lower dispersions between \textit{Gaia} DR3 and other catalogs. Thus, there are clear indications of a significantly different \textit{Gaia} DR3 stellar paramaters and abundances determined for the massive sequence stars compared to the RGB stars based on the comparison with the three external catalogs.

The calcium abundance trends as a function of metallicity for massive sequence stars in the Figures~\ref{fig:apg_gaia_validca_diff}, ~\ref{fig:galah_gaia_validca_diff}, and \ref{fig:cnn_gaia_validca_diff} show the \textit{Gaia} DR3 [Ca/Fe] distribution peaking at subsolar values with a large scatter while the distribution for the external catalogs peak at near solar values with smaller scatter. At the same time, the distributions for the RGB stars peak at solar value for both the \textit{Gaia} DR3 and the external catalogs with similar low scatter. We infer that the significant differences in the calcium abundances between \textit{Gaia} DR3 and other catalogs for the massive sequence stars could be attributed to the evident differences in \teff\, and \logg.

We confirm the absence of chemical depletion in the massive-star samples from both GALAH and APOGEE, with GALAH showing RGB-like magnesium and silicon abundances and APOGEE showing RGB-like silicon abundances, as illustrated in Fig.~\ref{fig:apogee_galah_mgsica}. However, APOGEE does display lower magnesium abundances that fall below the RGB sequence, a feature we investigate in detail in the next subsection. In summary, there is no evidence for chemical depletion in calcium, silicon, or magnesium among the massive sequence stars in the three external catalogs—except for magnesium in APOGEE.

 \begin{figure*}
  \includegraphics[width=\textwidth]{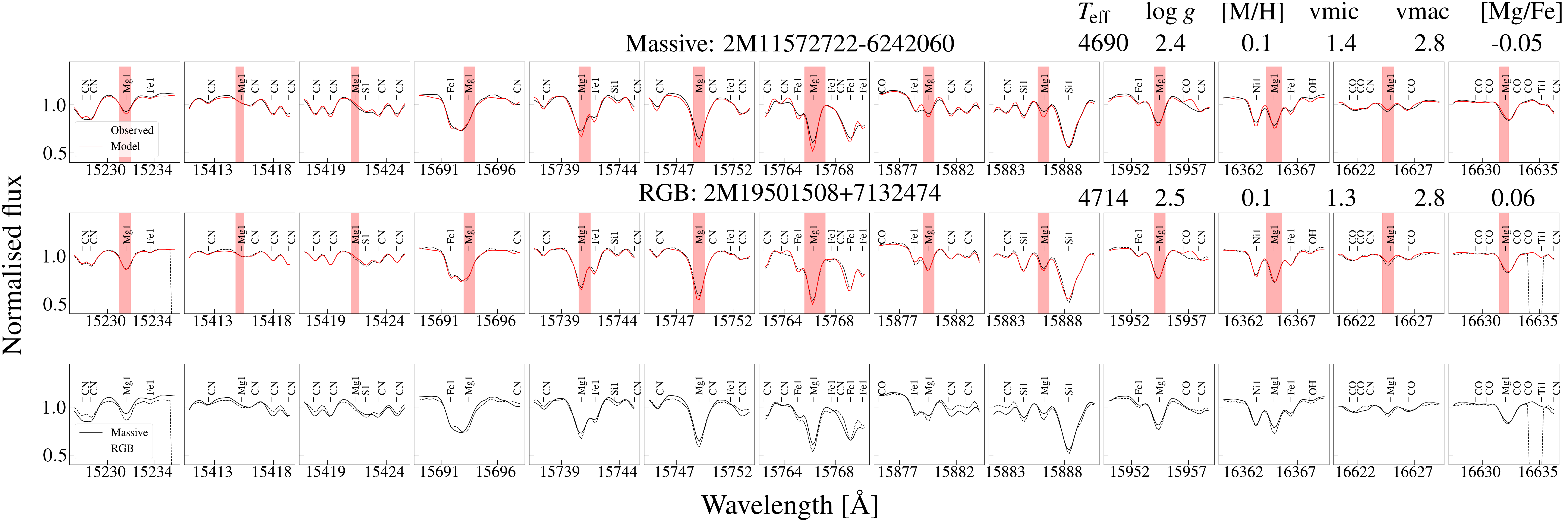}
  \caption{Spectral windows showing few Mg lines used for abundance determination in APOGEE. In each panel, APOGEE Observed (black) and ASPCAP model (red) spectra of one massive seuqnce star (solid lines) and one RGB star (dashed lines) are shown. The APOGEE$\_$ID, calibrated \teff, calibrated \logg, \feh\,, microturbulence, macroturbulence, and calibrated [Mg/Fe] values for each star are listed on top of the top two row panels.   }
  \label{fig:apogee_spectra_mg}
\end{figure*}

\subsection{Low magnesium massive sequence stars in APOGEE}
\label{sec:lowmg_apogee}

We expect the reasons, if any, for the inconsistent low magnesium abundances in the massive sequence stars in APOGEE to be evident when comparing the APOGEE observed spectra with and ASPCAP solution models. We identified 166 stars with low magnesium abundances among the 276 APOGEE massive sequence stars sample based on their location in the Figure~\ref{fig:apogee_galah_mgsica} (left top panel). Both the observed APOGEE spectrum and the ASPCAP model spectrum are available in the SDSS Science Archive Server \footnote{https://dr16.sdss.org/infrared/spectrum/view/}. In the \textit{aspcapStar} fits files, the pseudo-continuum normalised spectra are available.

In order to reject any effects in the spectral line features due to the differences in stellar parameters, one would need to look for massive-RGB star pairs with similar stellar parameters (\teff, \logg, \feh\,, and micro turbulence). However, due to the selection criteria applied to select massive and RGB stars, there is no significant overlap in stellar parameters between the two populations (see the bottom left panel of the Figure~\ref{fig:apg_gaia_diff}). Still, we could find very few massive sequence star-RGB star pairs with very similar stellar parameters.

In Figure~\ref{fig:apogee_spectra_mg}, we present the observed and ASPCAP model spectra for selected massive and RGB stars, focusing on the magnesium lines used in the APOGEE analysis. Each figure contains two rows: the first row shows the observed (black) and ASPCAP model (red) spectrum for a selected massive sequence star and RGB star, respectively. The third row displays the observed spectra of the massive (solid black line) and RGB (dashed black line) stars.
The APOGEE IDs or 2MASS IDs of the stars, along with their stellar parameters is also noted on top of each row. These parameters include the calibrated effective temperature (\teff), \logg, \feh\,, microturbulence, macroturbulence, and the  calibrated elemental abundance values determined in APOGEE. The stellar parameters are remarkably similar, with only slight differences in \teff\, (24 K), \logg\, (0.1 dex), and microturbulence (0.1 km s$^{-1}$). However, there are significant differences in the carbon (0.15 dex), nitrogen (0.08 dex), and oxygen (0.08 dex) abundances between the selected massive and RGB stars in addition to the obvious differences in masses and radii.
Similar figures showing the silicon and calcium lines used in the APOGEE abundance analysis can be found in Figures~\ref{fig:apogee_spectra_si} and \ref{fig:apogee_spectra_ca} in Appendix~\ref{sec:additional}.

From the Figure~\ref{fig:apogee_spectra_mg}, it is evident that the core of the majority of the selected magnesium lines are deep. Comparing the models to the observed spectra of the massive and RGB stars, we note that the models have several stronger Mg lines than the observed ones for the massive sequence star. For example, the models to the core of the magnesium lines at 15231 \AA, 15740 \AA, 15748 \AA, 15765 \AA, 15879 \AA, 15886 \AA, 15954 \AA, and 16365 \AA, in the massive sequence star are significantly stronger than the models to the same lines in the RGB star. We do not find a similar behaviour in the strength of the modelled silicon and calcium lines for massive sequence stars compared to the RGB stars (see the Figures~\ref{fig:apogee_spectra_si} and \ref{fig:apogee_spectra_ca}). Meanwhile, from the comparison of massive and RGB observed spectra, in the third row of the figures, we clearly see that the lines in the massive sequence star are broader than the ones of the RGB star (see for example, magnesium lines at 15740 \AA, 15879 \AA, 15886 \AA, silicon lines at 15375 \AA, 16163 \AA, 16681 \AA, 16828 \AA, and calcium lines at 16136 \AA, 16151 \AA, 16155 \AA, and 16157 \AA). The neighboring lines belonging to other species also show similar extra broadening in the massive sequence star spectrum compared to the RGB star spectrum. 

We note that the broadening parameters, microturbulence and macroturbulence, are similar for both  stars. Thus the model spectrum does not take into account the extra broadening in the massive sequence star spectrum, and this is possibly the reason for the increased strength in the model spectrum of the massive sequence star. Since the modelled lines are too deep, the abundance actually needs to be even lower which leads to lower magnesium abundances for massive sequence. Furthermore, while APOGEE determines the microturbulence velocity by setting it as a free parameter during the parameter determination step, the macroturbulence velocity is adopted for giants based on a parametric relation with metallicity \citep{jonsson:2020}. Moreover, the rotational parameter, $vsini$, is not included in the giant grids in APOGEE, based on the assumption that the giants do not have significant rotation. 
We note, however, as mentioned in the APOGEE DR17 website$\footnote{https://www.sdss4.org/dr17/irspec/parameters/}$, in the case of giants that have significant rotation, the ASPCAP results may be unreliable.

As a consequence of the points above, we speculate that the assumed macroturbulent velocity, set solely based on the metallicity of the star, is not able to model the extra broadening in the spectral lines of massive sequence stars. This seems to affect the modelling of stronger lines that could have more weight in the magnesium abundance estimation, leading to the determination of lower magnesium abundances for such stars. To further verify this, we carried out our own abundance analysis and redetermined magnesium, silicon, and calcium abundances from few selected lines for the 166 massive sequence stars and randomly selected 100 RGB stars with the macroturbulence velocity set as a free parameter (see appendix~\ref{sec:reanalysis}). A few examples illustrating the improved fit between the synthetic and observed spectra, obtained by allowing the macroturbulence velocity to vary freely compared to adopting the APOGEE macroturbulence values, are shown for five magnesium lines as well as across the full APOGEE spectral windows in the figures~\ref{fig:apogee_spectra_1}-\ref{fig:apogee_spectra_4} in the appendix~\ref{sec:additional}. Although we found increased magnesium abundances for massive sequence stars on changing the macroturbulence velocities, we note that we employ different linelists, NLTE grids, models, etc. for our analysis while adopting the APOGEE stellar parameters determined with the ASPCAP pipeline. Thus, in order to investigate this point more thoroughly (which is out of the scope of this paper), it would be ideal to carry out a detailed spectroscopic analysis starting from stellar parameter determination or carry out independent high resolution spectroscopic observations of low magnesium massive sequence stars in APOGEE. 

Meanwhile, GALAH determines the rotational velocity, $vsini$, during parameter determination for their stars and thus takes into account the line broadening with this parameter. We find a clear increasing trend of $vsini$ with increase in masses for the massive sequence stars in the Figure~\ref{fig:Galah_mass_vsini} which is not evident in the case of RGB stars. This supports our speculation of broader absorption lines in massive sequence stars, which has been taken into account in GALAH while determining abundances.

Another possibility is that the low magnesium in massive sequence stars maybe due to possible NLTE related issues in the modelling of magnesium lines. We note that there is significant difference in the \teff\,(upto 1000 K) between the massive and RGB stars samples (see the APOGEE Kiel diagram in the Figure~\ref{fig:apg_gaia_diff}). In case the NLTE models used in the APOGEE analysis have a \teff\,-dependence (for e.g. see Mangnese trends in the Figure 15 in \cite{Nandakumar:24_21elements}), this could affect the magnesium abundance determination from the stronger magnesium lines, and lead to the discrepancy that is seen only for magnesium. Again, detailed high resolution spectroscopic analysis  will be needed to investigate this scenario.

\subsection{Implication on the chemical evolution model}
 
\citet{Spitoni:2023}  extended the classical two-infall model \citep{Chiappini:1997, Spitoni:2019}  by introducing an additional gas accretion episode during the thin-disc phase. This extra infall, occurring within the last ~2 Gyr of Galactic evolution, was motivated by the recent peaks in the star-formation history (SFH)   by \citet{Ruiz-Lara:2020} and interpreted as a consequence of the interaction between the Milky Way and the Sagittarius dwarf galaxy.

Within that framework, \citet{Spitoni:2023} succeeded in reproducing both the SFH of \citet{Ruiz-Lara:2020} and the population of massive sequence stars revealed by the Gaia DR3 analysis, which suggested the presence of chemically depleted young stars.
 A key assumption of this model is that roughly 30\% of the gas accreted onto the thin disc, with primordial chemical composition, takes part in the most recent infall episode, i.e. the remaining 70\% is accreted during the first thin-disc infall phase.
This led to a significant dilution signature in the [Si/Fe]–[Fe/H] plane, as shown in Fig. \ref{fig:CEM}.

In the same work, it was also demonstrated that a chemical evolution model characterised by similar SFH and the same total accreted mass for the thin disc can nonetheless produce different chemical evolution paths if the infalling gas is distributed differently among the various accretion episodes. In particular, Fig. \ref{fig:CEM} shows a model in which the mass involved in the third infall episode is reduced to $15\%$ of the total thin-disc accretion while adopting a slightly higher star formation efficiency than that assumed by \citet{Spitoni:2023} for the most recent infall event (the “weak-dilution” case). Here the remaining 85\% is accreted during the first thin-disc infall phase. In this configuration, the dilution effect is considerably weaker, and the resulting track in the [Si/Fe]–[Fe /H] diagram closely resembles that predicted by the classical two-infall models of  \citet{Spitoni:2019,Spitoni:2020,Spitoni:2021}.

As highlighted in the present study, the reanalysis of the massive-star population does not appear to support the strong chemical dilution originally inferred from Gaia DR3.  Nevertheless, the three-infall scenario remains a plausible description of the formation history of the local Galactic disc. Moreover, the weak-dilution configuration is fully consistent with the three-infall model recently proposed by \citet{Palla:2024} to explain the abundance gradients traced by young open clusters in the Gaia-ESO survey.

\section{Conclusions}
\label{sec:conclusions}
\vspace{-5pt}

In this work, we carried out a detailed investigation on the massive sequence stars with low calcium abundances identified in the \textit{Gaia}-GSP Spec catalog \citep{Recio-Blanco:2023chemcart} with the objective to confirm their existence in other Milky Way spectroscopic survey catalogs. We applied the same selection criteria (with uncalibrated \teff\, and \logg) to select the massive and RGB stars sample from the \textit{Gaia}-GSP Spec catalog, and cross matched it with three external catalogs- APOGEE, GALAH, and \textit{Gaia}-CNN, with the objective to probe the $\alpha$ abundances in the same stars in other Milky Way spectroscopic surveys. We found significant number of stars (1000-15000) in common with the external catalogs in the case of RGB stars, and lower numbers (300-600) in the case of massive sequence stars (see Table~\ref{table:statistics}). We computed the masses of the selected sets of stars with an updated version of \cite{Kordopatis:2023a} by projecting the spectroscopic \teff, \logg, metallicity and the de-redenned infrared absolute magnitudes derived from 2MASS (J, H, Ks) and WISE (W1, W2) on the PARSEC isochrones. The estimated masses were validated with the masses from APOKASC3 and TESS.

Our initial comparison of the stellar parameters and calcium abundances between \textit{Gaia} DR3 and the three external catalog values revealed significant differences for massive sequence stars and better agreement for RGB stars (see figures~\ref{fig:apg_gaia_diff}, ~\ref{fig:galah_gaia_diff}, and ~\ref{fig:cnn_gaia_diff} and table~\ref{table:parameterdiff}). With the [Ca/Fe] ([$\alpha$/Fe] in the case of \textit{Gaia} CNN) versus [M/H] plots (figures~\ref{fig:apg_gaia_validca_diff}, ~\ref{fig:galah_gaia_validca_diff}, and ~\ref{fig:cnn_gaia_validca_diff}), we demonstrated the absence of the low calcium abundance massive sequence stars and its similarity with the RGB trends using the three external catalogs even though the low calcium abundance massive sequence stars are present in the \textit{Gaia} DR3 sample. We attribute this discrepancy to the significant difference in the stellar parameters of the massive sequence stars between \textit{Gaia} DR3 and the other catalogs, which in turn could result in low calcium abundances in \textit{Gaia} DR3. We also note that the use of \teff-based calibrations for calcium still resulted in low calcium abundances for the massive sequence stars from \textit{Gaia} DR3 in this work. Our probe into the strengths of calcium lines in the \textit{Gaia} spectra of massive sequence stars and RGB stars revealed differences that could be explained by the \teff\, dependence of these lines (see Figure~\ref{fig:gaiaspectra_ca}). Investigation of the other alpha species such as magnesium and silicon in the massive and RGB stars in GALAH and APOGEE revealed similar silicon abundances for the two populations in both APOGEE and GALAH. Magnesium abundances, however, was found to be lower for massive sequence stars with metallicities above -0.5 dex in APOGEE but not in GALAH. We speculate that the reason for the low abundances only in the case of magnesium in APOGEE could be due to the extra broadness in the absorption lines of massive sequence stars possibly due to rotation, combined with the APOGEE macroturbulence velocities for giants being set solely based on metallicity under the assumption that giants exhibit negligible rotation. This may not be the correct assumption for the young massive giant population, and could lead to wrong abundances and even ASPCAP stellar parameters. Alternatively, it could be due to the NLTE related issues affecting the stronger magnesium lines that have more weightage in abundance determination.

Based on our investigation, we conclude that the chemical depletion seen in the massive young stellar population using calcium abundances from \textit{Gaia} DR3 catalog is not evident in the calcium abundances determined for the common stars in APOGEE, GALAH, and \textit{Gaia} CNN. This is also evident from silicon in APOGEE and GALAH, and magnesium in GALAH. We speculate that the low magnesium abundances seen in a subset of APOGEE massive sequence stars could be an artifact due to the assumption of wrong broadness parameter for these stars. Further investigations with dedicated observations using high resolution spectroscopic instruments will be needed to confirm this. Thus alpha-abundances of massive sequence stars derived from the Gaia-RVS spectra should be used with  caution. However, the already proposed  three-infall chemical evolution model may not be discarded as the case with weak dilution could still explain the recent star formation episodes even with the absence of the massive young stellar population with chemically depleted signatures.

\begin{acknowledgements}
 We thank the referee for the valuable comments and suggestions that improved the quality of the paper. G.N. gratefully acknowledges the support from the Crafoord Foundation via the Royal Swedish Academy of Sciences (Vetenskapsakademiens stiftelser; CR 2024-0034). G.N. and M.S. thank Prof. Shashikiran Ganesh for hosting them at PRL, India where the discussions on this project commenced. G.N. also thanks the ISA group with Lund and Malmo university members for helpful discussions during weekly meetings.
 GK gratefully
acknowledges support from the French national research agency (ANR) funded
project MWDisc (ANR-20-CE31-0004). 
N.R.\ acknowledges support from the Swedish Research Council (grant 2023-04744) and the Royal Physiographic Society in Lund through the Stiftelsen Walter Gyllenbergs and Märta och Erik Holmbergs donations. 
E.S.and F.M.   thank I.N.A.F. for the  
1.05.24.07.02 Mini Grant - LEGARE "Linking the chemical Evolution of Galactic discs AcRoss diversE scales: from the thin disc to the nuclear stellar disc" (PI E. Spitoni). 

The following software and programming languages made this
research possible: TOPCAT (version 4.6; \citealt{topcat}); Python (version 3.8) and its packages ASTROPY (version 5.0; \citealt{astropy}), SCIPY \citep{scipy}, MATPLOTLIB \citep{matplotlib} and NUMPY \citep{numpy}.
\end{acknowledgements}

%
%


\bibliographystyle{aa}
\bibliography{references} 



\appendix

\begin{appendix} 

\section{Comparison with \textit{Gaia} CNN}
\label{sec:gaiacnn}

\begin{figure*}[htp]
\centering
  \includegraphics[width=0.8\textwidth]{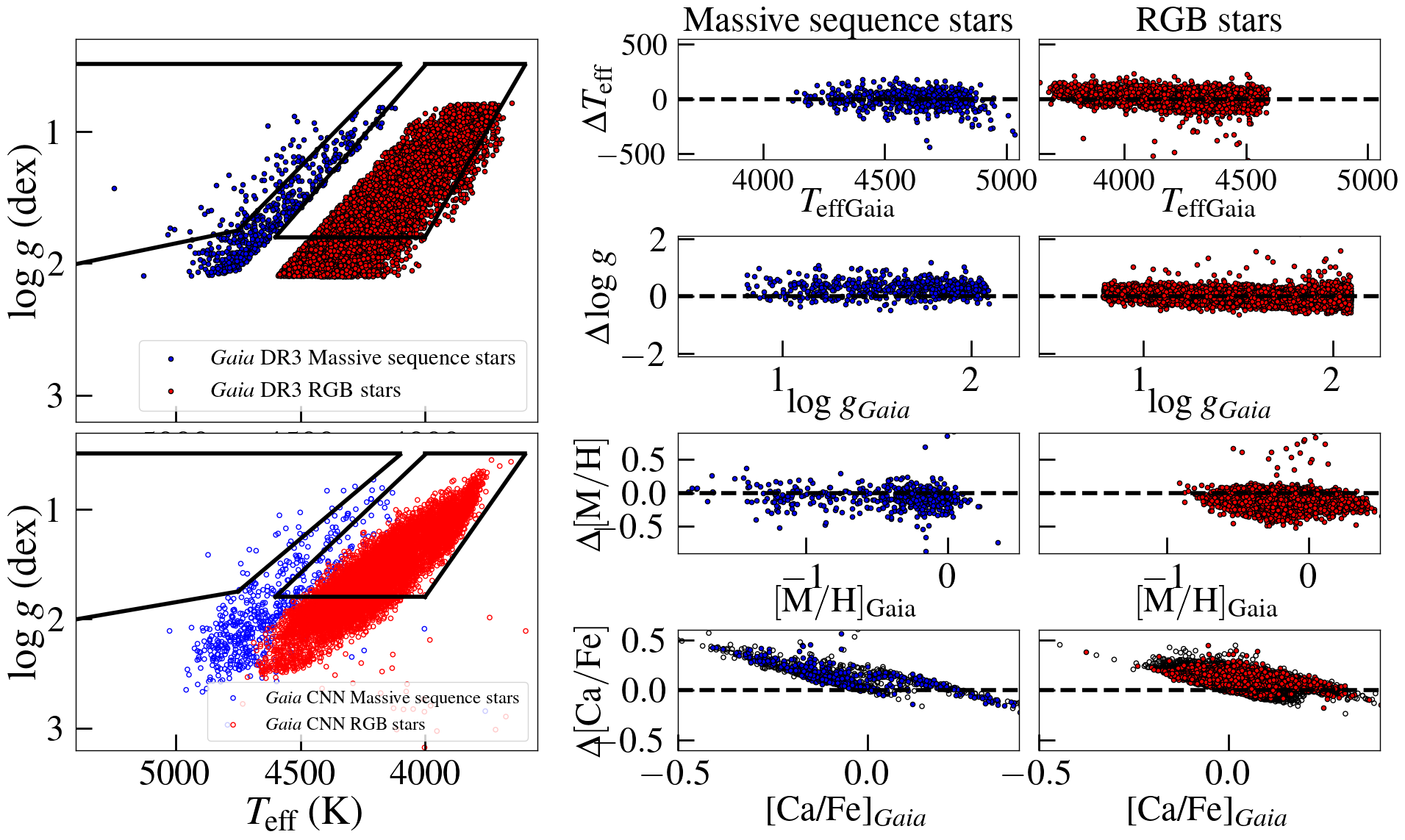}
  \caption{ Similar to the Figure~\ref{fig:apg_gaia_diff}, but showing the comparison of the stellar parameters and calcium abundances between the massive (blue) and RGB (red) stars in common between \textit{Gaia}-CNN and \textit{Gaia} DR3. The \textit{Gaia} DR3 [Ca/Fe] is compared with the general apha abundances in \textit{Gaia} CNN in the plots in the fourth row middle and last column panels.}
  \label{fig:cnn_gaia_diff}%
\end{figure*}

\begin{figure*}[htp]
 \centering
  \includegraphics[width=0.8\textwidth]{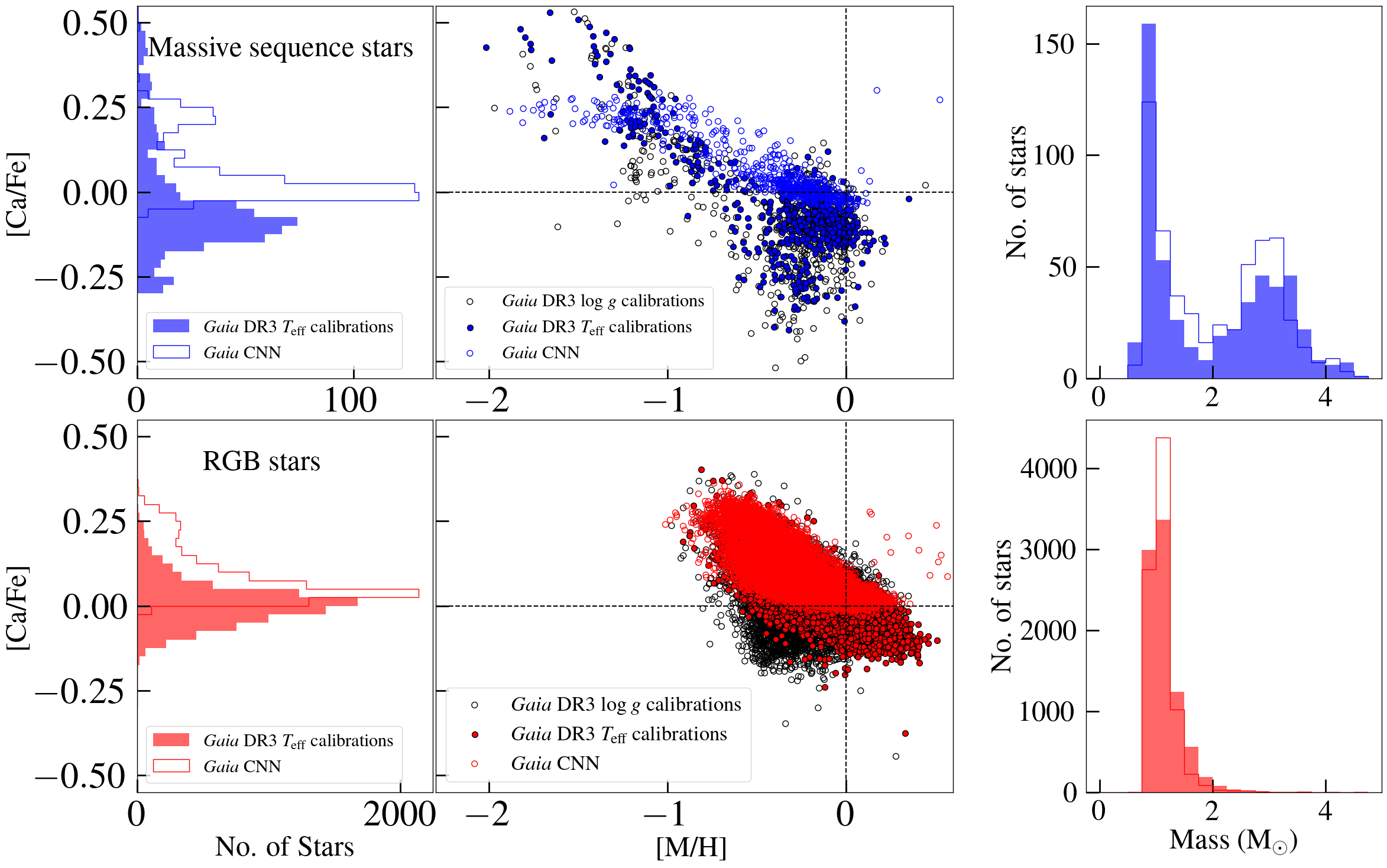}
  \caption{ Similar to the Figure~\ref{fig:apg_gaia_validca_diff}, but for the common stars in \textit{Gaia}-CNN and \textit{Gaia} DR3.}
  \label{fig:cnn_gaia_validca_diff}%
\end{figure*}

The cross-match of the \textit{Gaia} CNN catalog with the \textit{Gaia} DR3 massive and RGB stars samples with valid calcium abundances resulted in 573 stars and 8561 stars respectively. We did not apply any further filter to the \textit{Gaia} CNN catalog. 
The Kiel diagrams and the differences in stellar parameters and calcium abundances for massive and RGB stars in common between \textit{Gaia} DR3 and \textit{Gaia} CNN are shown in the Figure~\ref{fig:cnn_gaia_diff}, with the mean differences and standard deviations listed in the Table~\ref{table:parameterdiff_cnn}. \textit{Gaia} CNN only provides the general alpha abundances which we used to compare with \textit{Gaia} DR3 calcium abundances in the Figure~\ref{fig:cnn_gaia_diff}. The Figure~\ref{fig:cnn_gaia_validca_diff} shows the [Ca/Fe] versus [M/H] trends, [Ca/Fe] distributions, and mass distributions for the common massive and RGB stars.

There is better agreement in \teff  and \logg  for massive sequence stars between \textit{Gaia} DR3 and \textit{Gaia} CNN compared to APOGEE and GALAH (see Section~\ref{sec:comparison_stellarparameters}), but the dispersion in still high for both the parameters. The mean differences and dispersions in \feh\, and [Ca/Fe] are similar to APOGEE and GALAH. The difference trend in the case of calcium abundances is similar to APOGEE but with more stars. This is expected as the \textit{Gaia} CNN alpha abundances are predicted using CNN models trained on the APOGEE values. Similar to APOGEE, the \textit{Gaia} CNN [Ca/Fe] distribution for massive sequence stars peaks at solar value, with the distribution skewed towards super solar values, and the [Ca/Fe] versus [M/H] trends are tight with no indication of depletion. The mass distribution shows two peaks, close to 1 M$_{\odot}$ and 3 M$_{\odot}$, with no significant shift in the peaks compared to \textit{Gaia} DR3 masses.

For the RGB  stars, the mean differences and dispersions in the stellar parameters and calcium abundances are similar to APOGEE and GALAH. The [Ca/Fe] distribution and [Ca/Fe] versus [M/H] trends of \textit{Gaia} CNN RGB stars are tighter and are constrained to solar and above abundance values, which is not the case of APOGEE. The mass distribution is very similar to \textit{Gaia} DR3, APOGEE, and GALAH with masses less than 2 M$_{\odot}$.

\begin{table}
\small
\caption{ The mean and standard deviation of the differences in \teff, \logg, \feh\,, and [Ca/Fe] between the values in \textit{Gaia} DR3 and \textit{Gaia} CNN for the respective masssive and RGB star samples. The values in brackets are the mean and standard deviation in calcium abundances when using \logg\, based calibrations for \textit{Gaia}.}\label{table:parameterdiff_cnn}
\begin{tabular}{|c | c | c c c c |}
 \hline
Catalog &  & $\Delta$\teff & $\Delta$\logg & $\Delta$\feh\, & $\Delta$[Ca/Fe]    \\
\hline 
\multirow{4}{*}{\textit{Gaia} CNN}   & $\mu_{massive}$ & -11 & 0.29 & -0.07 & 0.11 (0.14) \\
& $\sigma_{massive}$ &  127 & 0.26 & 0.20 & 0.12 (0.13) \\
& $\mu_{RGB}$ & 18 & 0.02 & -0.09 &  0.08 (0.06) \\
& $\sigma_{RGB}$ & 60 & 0.18 & 0.14 & 0.05 (0.07) \\ 
\hline
\end{tabular}
\end{table}

\section{Representativeness of the cross-matched subsets in APOGEE and GALAH}
\label{sec:representativeness}

\begin{figure*}[htp]
\centering
  \includegraphics[width=0.9\textwidth]{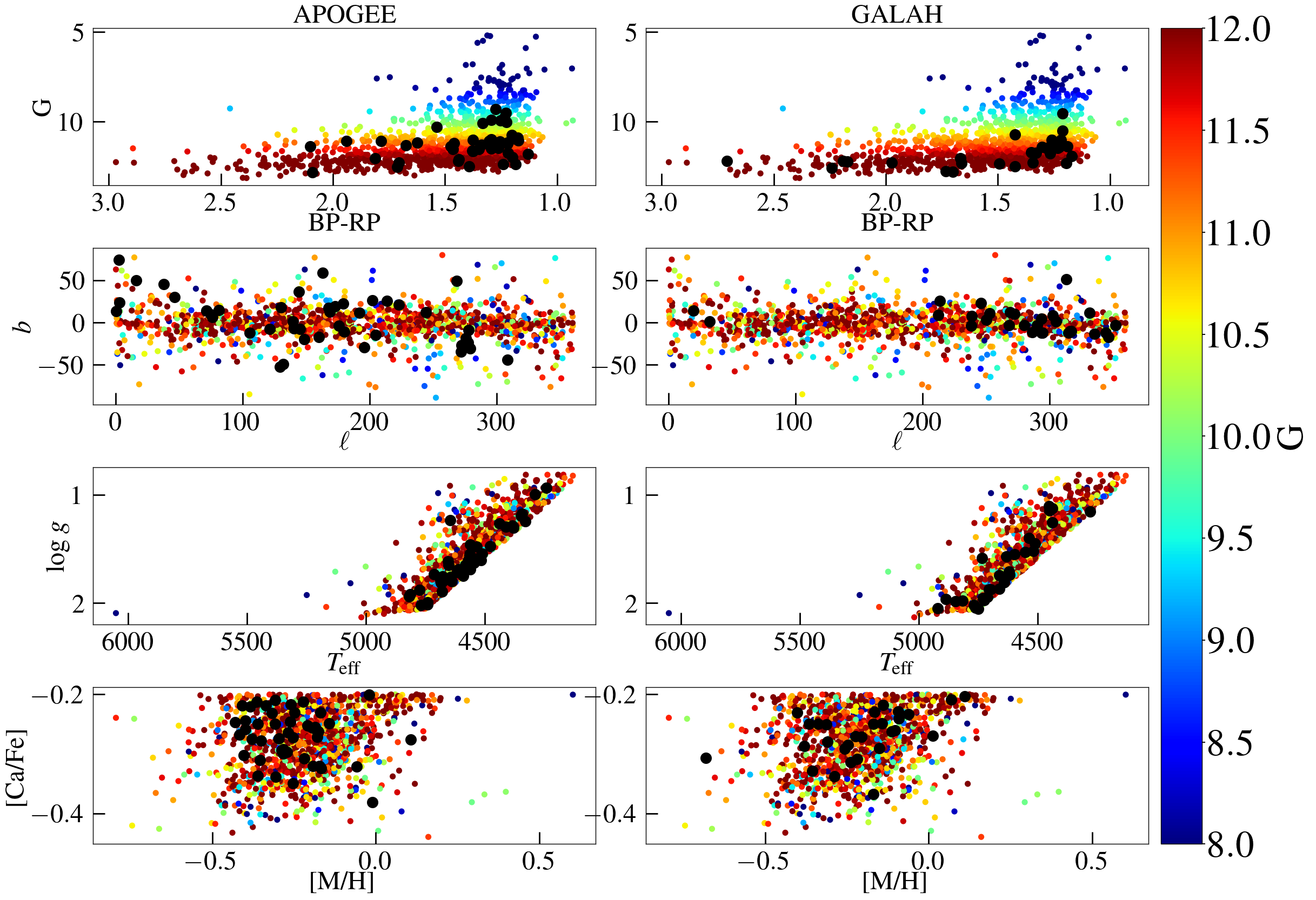}
  \caption{Parameter coverage of low calcium (\textit{Gaia}[Ca/Fe]<-0.2 dex) stars in \textit{Gaia} (colored circles) that are also observed in APOGEE (left; black filled circles) and GALAH (right; black filled circles). The first to fourth rows represent color magnitude digram (G vs BP-RP), spatial coverage ($b$ vs $\ell$), Kiel diagram (\teff\, vs \logg), and calcium abundance trend ([Ca/Fe] vs [M/H]) respectively, with the \textit{Gaia} stars color coded with their G magnitudes.   }
  \label{fig:selection}
\end{figure*}

The relatively low number of massive sequence stars ($\sim$300) in the APOGEE and GALAH cross matched catalogs raises a concern that these cross-matched stars may not adequately sample the low-[Ca/Fe] massive sequence stars ($\sim$16800) in \textit{Gaia}. To investigate the representativeness of the cross-matched subsets in APOGEE and GALAH, in the figure~\ref{fig:selection} we plot the color magnitude digram (G vs BP-RP; first row), spatial coverage ( $b$ vs $\ell$; second row), Kiel diagram (\teff\, vs \logg\, with \textit{Gaia} values; third row), and calcium abundance trend ([Ca/Fe] vs [M/H] with \textit{Gaia} values; fourth row) for the low calcium (\textit{Gaia} [Ca/Fe]<-0.2 dex) stars in \textit{Gaia} (colored circles) with the common stars in APOGEE (left) and GALAH (right) shown in black filled circles. 

From the [Ca/Fe] vs [M/H] plots, it can be seen that the common stars in APOGEE and GALAH cover the [Ca/Fe] and [M/H] range spanned by the low [Ca/Fe] stars in the \textit{Gaia} sample, and that the lowest [Ca/Fe] stars in the Gaia sample do not extend much beyond -0.4 dex. From the color magnitude diagrams, it is evident that the common stars in APOGEE and GALAH span the color range covered by \textit{Gaia} low [Ca/Fe] stars, but stars brighter than G$\sim$9 mag are not present in both APOGEE and GALAH samples which may incur a selection bias. At the same time, we note that the brighter stars (G $<$ 9 mag) do not preferentially clump together at the lowest [Ca/Fe] (see the [Ca/Fe] vs [M/H] panels). The lowest [Ca/Fe] stars are a mix of faint and bright stars, and dominated largely by faint stars as also evident from the CMD. From the \teff\, vs \logg\, diagrams, it can be seen that there is significant overlap of APOGEE/GALAH stars with \textit{Gaia} stars, and the low [Ca/Fe] stars have \textit{Gaia} Teff <5000 K. In terms of spatial coverage, there is a gap in the spatial location of stars in GALAH, missing stars within 40$^{o}$ $<$ $\ell$ $<$ 200$^{o}$. This gap is simply a consequence of the observing location limitations from the South Hemisphere. But there is no significant bias in the spatial location of stars in APOGEE. 
 
Overall, we show that even though the numbers are small in APOGEE and GALAH, both cross match samples cover similar \teff-\logg, G vs BP-RP parameter spaces as done by \textit{Gaia} stars. The brighter stars are missing in APOGEE and GALAH, but we show that there is no preferential magnitude range for low [Ca/Fe] stars. The lack of overlap in ($\ell$,$b$) range for GALAH should not have any impact on our results since both the inner and outer disc regions are still covered.

\section{Re-analysis of APOGEE spectra}
\label{sec:reanalysis}

\begin{figure*}[htp]
 \centering
  \includegraphics[width=0.9\textwidth]{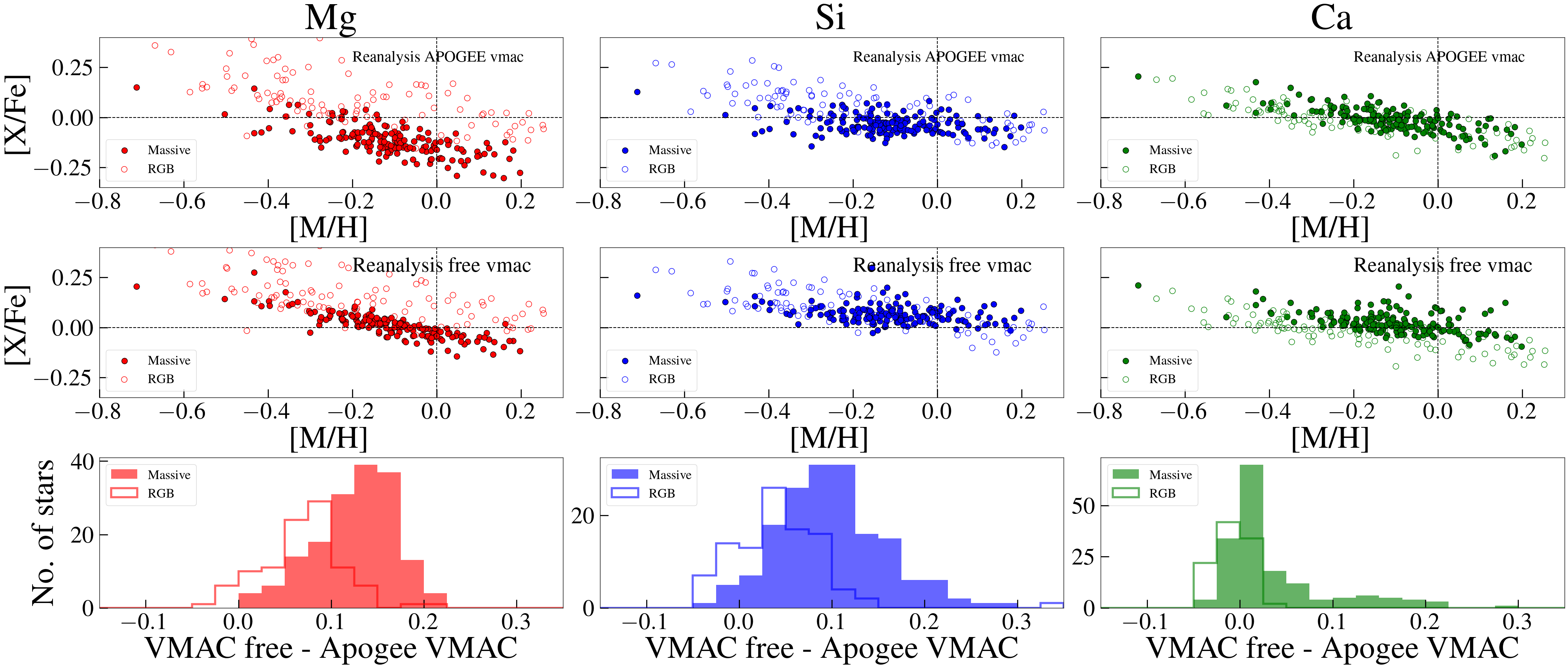}
  \caption{ Abundance trends as a function of metallicity for magnesium (left panel), silicon (middle panel), and calcium (right panel) for the selected 166 massive (filled circles) stars and 100 RGB (open circles) stars redetermined from selected few lines in this work with macroturbulence velocities adopted from APOGEE in the top row and those determined with macroturbulence velocities set as free paramter in the middle row. The bottom row shows the distribution of the abundances difference between when determined with determined when macroturbulence velocities are set free and when APOGEE macroturbulence velocities are adopted for massive (filled histograms) and RGB (open histograms) stars.}
  \label{fig:reanalysis}
\end{figure*}

As mentioned in Section~\ref{sec:lowmg_apogee}, we carried out our own abundance analysis to redetermine magnesium, silicon, and calcium abundances for 166 massive sequence stars and randomly selected 100 RGB stars. For our analysis, we used the spectroscopy made easy (SME) tool along with MARCS stellar atmosphere models and an updated linelist (see \citealt{Nandakumar:2023,Nandakumar:24_21elements} for more details). We used APOGEE stellar parameters and determined abundances for two cases: (i) macroturbulence velocity set to APOGEE values and (ii) macroturbulence velocity set as a free parameter. In the top two rows of the figure~\ref{fig:reanalysis}, we show the redetermine abundance trends as a function of metallicity for magnesium (left panel), silicon (middle panel), and calcium (right panel) for the selected 166 massive (filled circles) stars and 100 RGB (open circles) stars. The top and middle rows show the abundances determined with macroturbulence velocities adopted from APOGEE and those determined with macroturbulence velocities set as free paramter respectively. The bottom row shows the distribution of the abundances difference between when determined with determined when macroturbulence velocities are set free and when APOGEE macroturbulence velocities are adopted for massive (filled histograms) and RGB (open histograms) stars.
We find higher magnesium abundances when macroturbulence velocity is set free for massive sequence stars ($\sim$0.15 dex) and a much narrower trend compared to when using APOGEE mactroturbulence values. Slight increase in magnesium abundances ($<$0.1 dex) is seen in the case of RGB stars as well, placing the massive sequence stars still below the RGB stars in the [Mg/Fe] versus [M/H] plot. Silicon abundances demonstrate comparable increase for both massive and RGB stars when macroturbulence velocity is set free leading to the overlap in their abundance trends. Surprisingly, calcium abundances for RGB stars lie below the massive sequence stars upon setting the macroturbulence velocity free. We note that we adopt the APOGEE stellar parameters determined with the ASPCAP pipeline and then we employ different linelists, NLTE grids, models etc for our analysis. This is not ideal since the wrong macroturbulence velocities could lead to wrong stellar parameters which we didn't take into account for our analysis.

\section{Additional figures}
\label{sec:additional}

\begin{figure*}[htp]
\centering
  \includegraphics[width=0.9\textwidth]{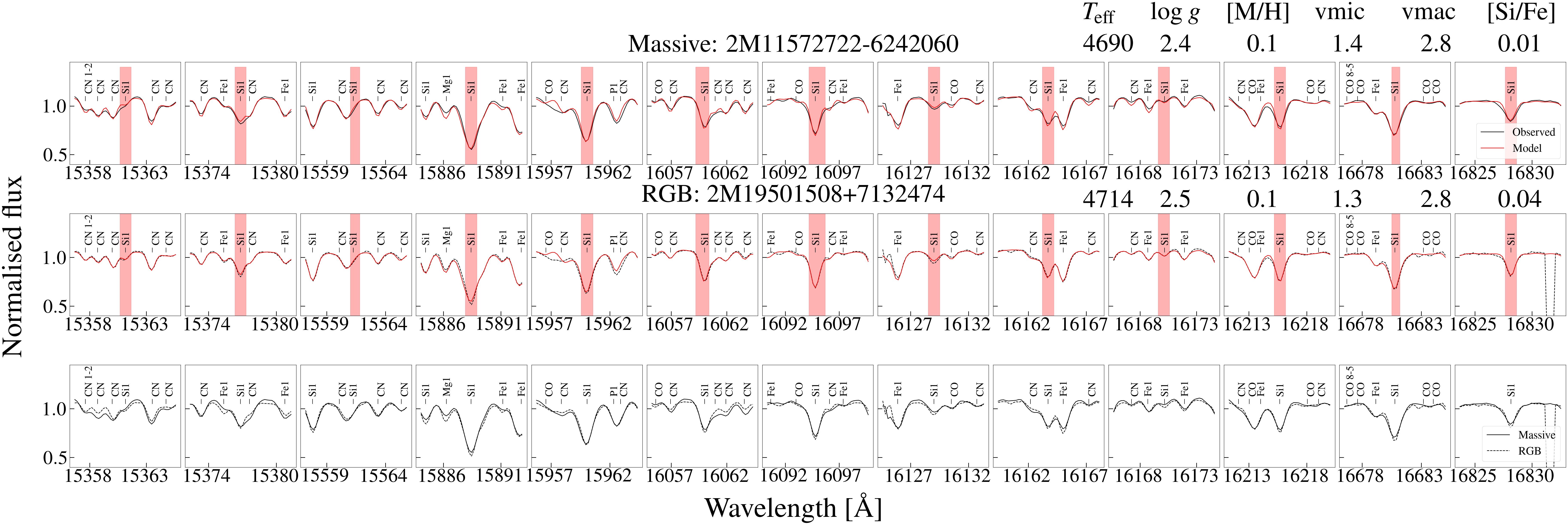}
  \caption{Similar to the Figure~\ref{fig:apogee_spectra_mg} but for silicon.   }
  \label{fig:apogee_spectra_si}
\end{figure*}

 \begin{figure*}[htp]
 \centering
  \includegraphics[width=0.9\textwidth]{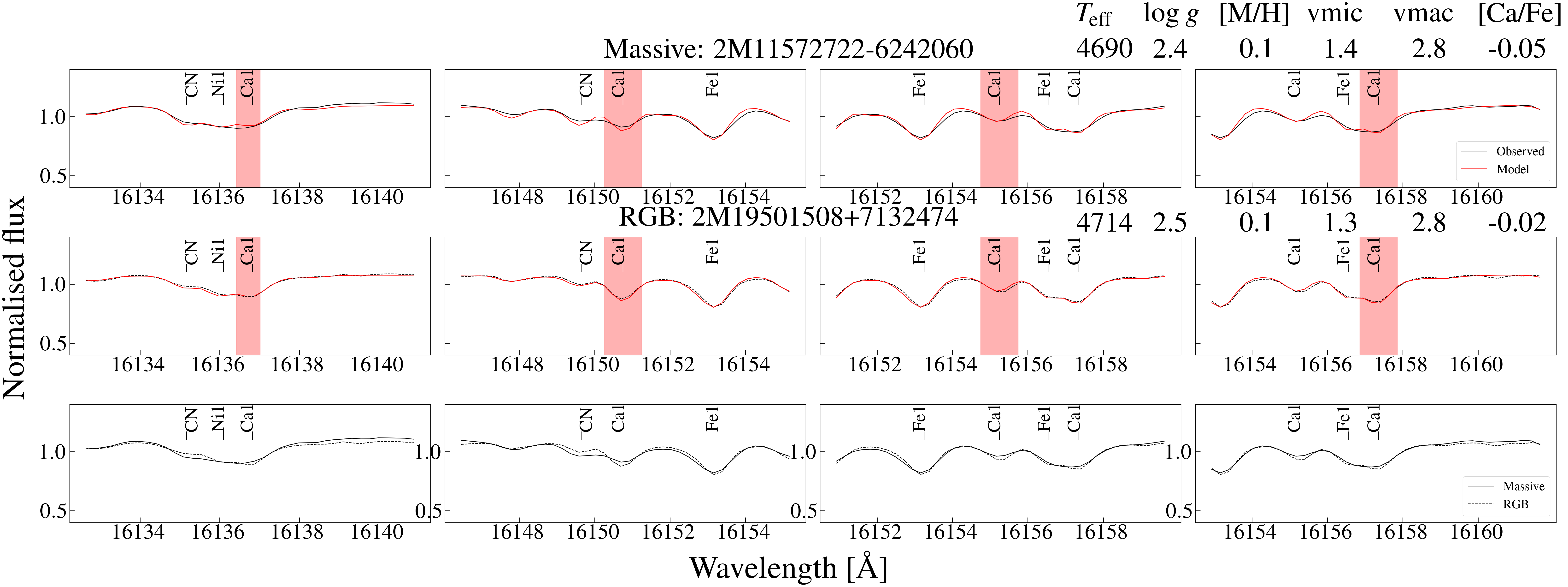}
  \caption{Similar to the Figure~\ref{fig:apogee_spectra_mg} but for calcium.   }
  \label{fig:apogee_spectra_ca}
\end{figure*}

\begin{figure*}[t]
\centering

\begin{minipage}{0.48\textwidth}
    \centering
    \includegraphics[width=\linewidth]{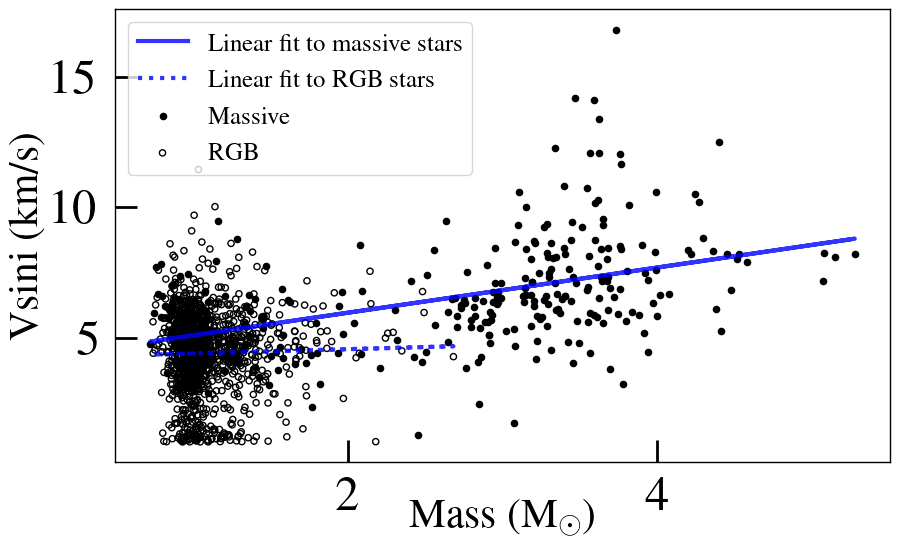}
    \captionof{figure}{Mass versus $vsini$ trend for massive (filled black circles) and RGB (open black circles) stars used in this work from the GALAH DR4 catalog. Stellar masses are estimated by projecting their atmospheric parameters from GALAH DR4 and infrared magnitudes onto PARSEC isochrones. $vsini$ is determined in the GALAH analysis pipeline. The linear fit to the massive (solid blue line) and RGB stars (dotted blue line) shows a clear increasing trend in $vsini$ for higher mass stars which is not evident in the low mass RGB stars.}
    \label{fig:Galah_mass_vsini}
\end{minipage}
\hfill
\begin{minipage}{0.48\textwidth}
    \centering
    \includegraphics[width=\linewidth]{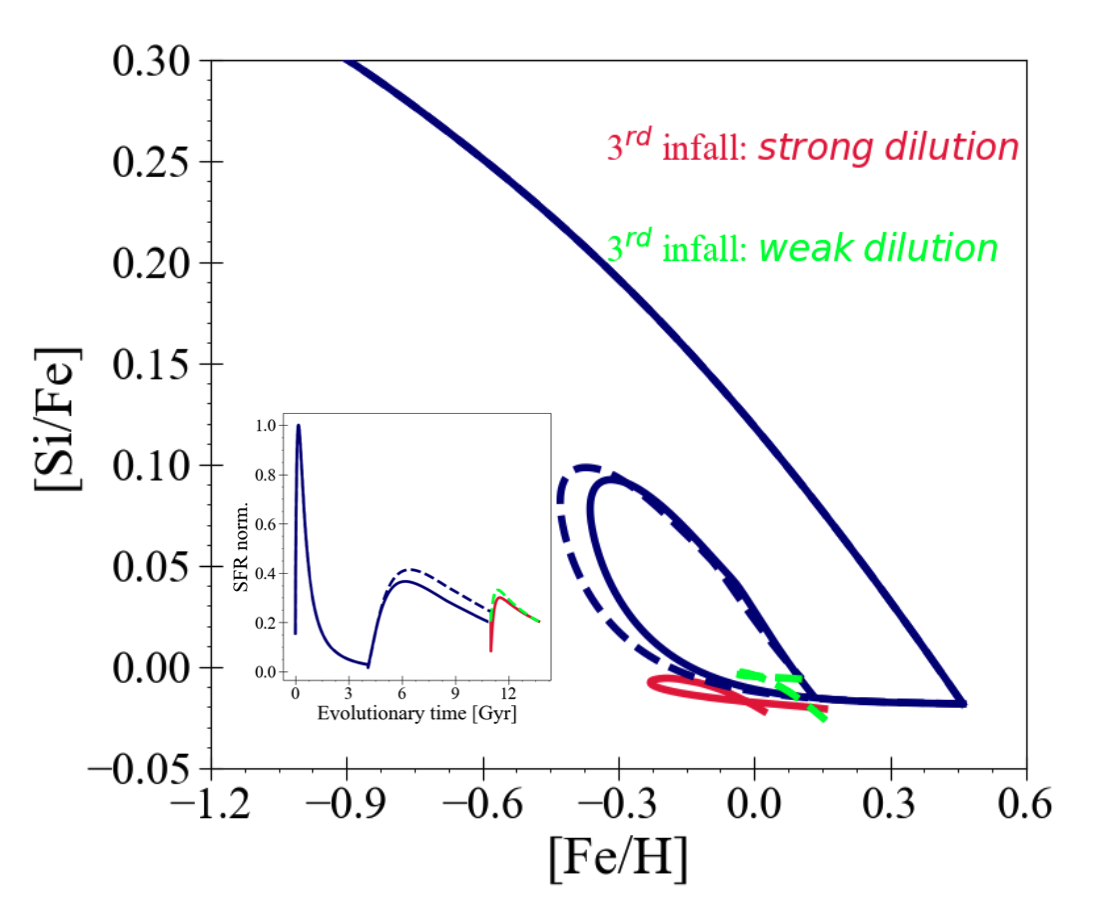}
    \captionof{figure}{Comparison between three-infall chemical evolution models with different levels of dilution during the thin-disc phase.
The solid line represents the model of \citet{Spitoni:2023}, in which the gas accreted during the third infall accounts for 30\% of the total baryonic mass of the thin disc (strong dilution case; red segment),  i.e. the remaining 70\% is accreted during the first thin-disc infall phase. The dashed line shows the model where the third infall contributes only 15\% of the thin-disc mass (weak dilution case; green segment). The inset additionally shows the normalized star formation history for  the various evolutionary phases in the two presented models.}
    \label{fig:CEM}
\end{minipage}

\end{figure*}

 \begin{figure*}[]
 \centering
  \includegraphics[width=0.9\textwidth]{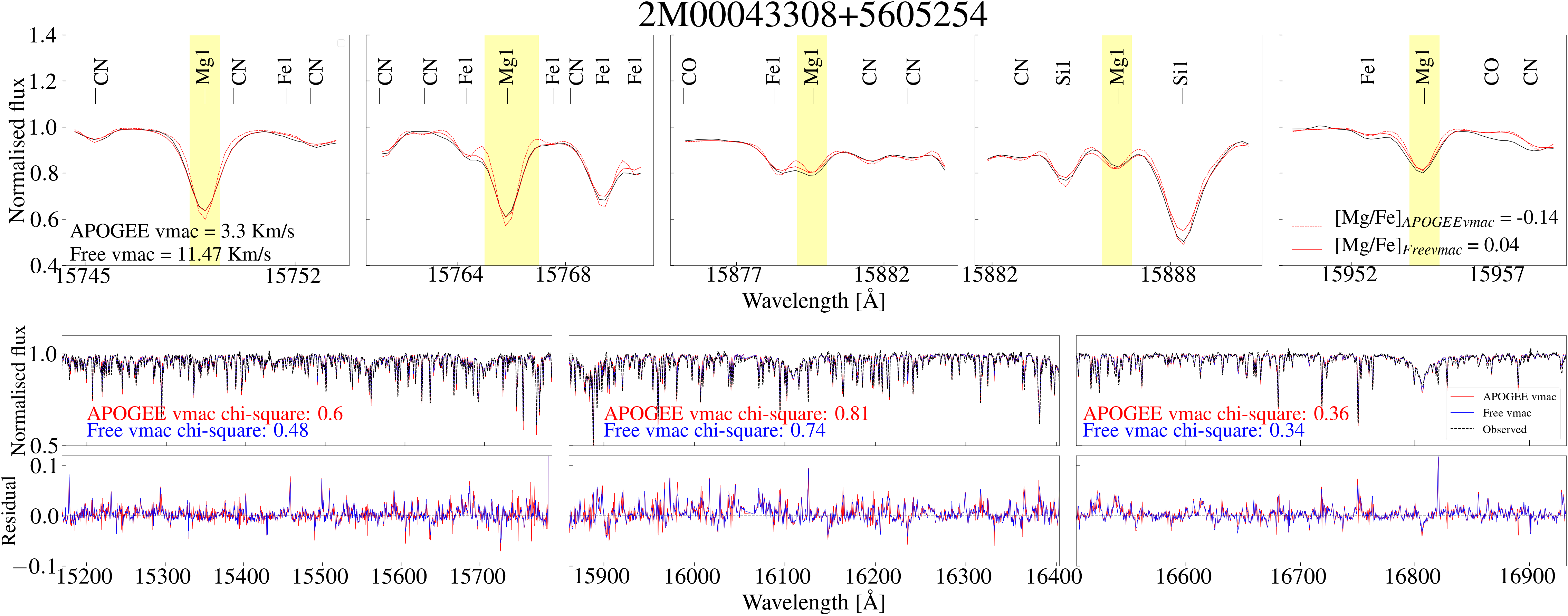}
  \caption{Reanalysis of APOGEE spectrum for the massive sequence star 2M00043308+5605254 setting macroturbulence (vmac) free and using APOGEE vmac, while keeping all other stellar parameters as provided in APOGEE DR17. The spectral windows showing the five magnesium lines used in the abundance reanalysis is shown in the top row with the black line representing the observed star spectrum, the solid red line representing the synthetic spectrum generated using APOGEE vmac, and the dashed red line representing the synthetic spectrum generated on setting vmac free. Both vmac values and the resulting magnesium abundances are also listed in the plot. In the middle row, the full wavelength range (three chip segments) of APOGEE observed spectrum is shown in black dotted line, overlaid with the synthetic spectra generated using APOGEE vmac (red line) and on setting vmac free (blue line). The difference in flux between the observed spectrum and synthetic spectra generated using APOGEE vmac (red line) and on setting vmac free (blue line) are shown in the bottom row. The chi-square value of the fits of both synthetic spectra to the observed spectrum in each chip is estimated and listed in the middle row.}  
  \label{fig:apogee_spectra_1}
\end{figure*}

\begin{figure*}[]
 \centering
  \includegraphics[width=0.9\textwidth]{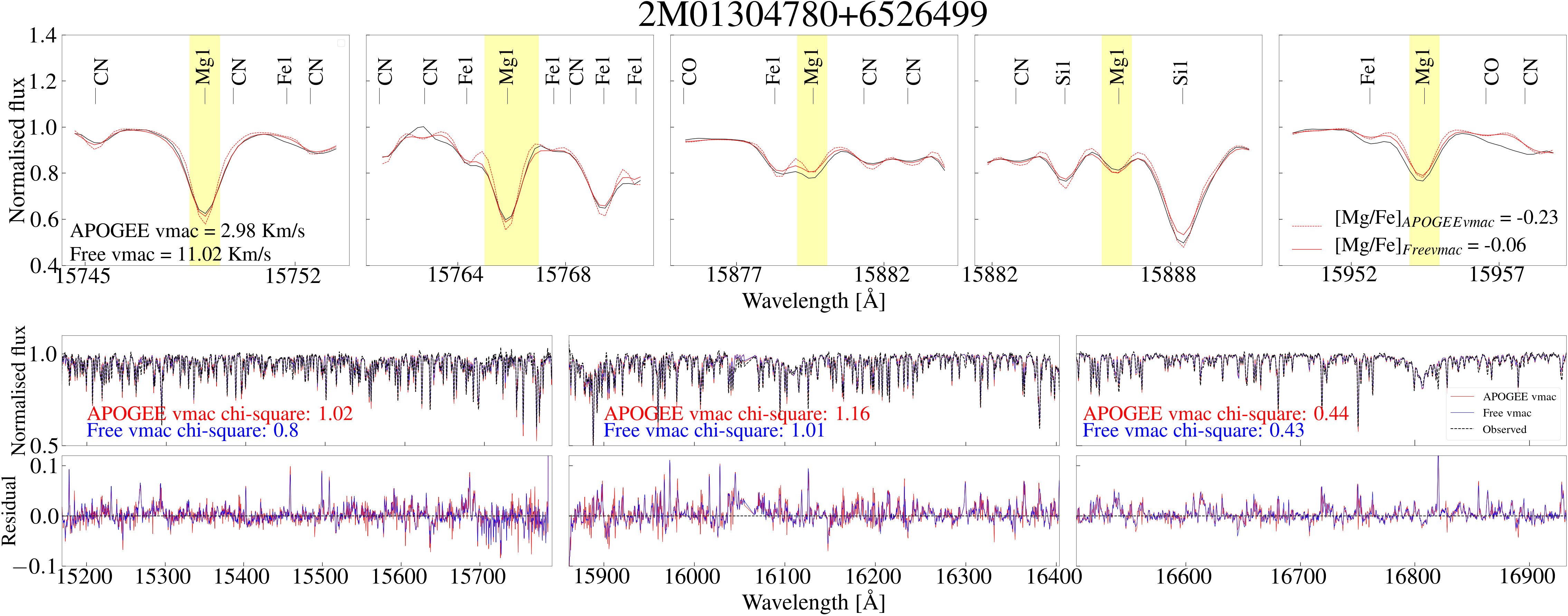}
  \caption{ Same as figure~\ref{fig:apogee_spectra_1}, for the massive sequence star 2M01304780+6526499. }
  \label{fig:apogee_spectra_2}
\end{figure*}

\begin{figure*}[]
 \centering
  \includegraphics[width=0.9\textwidth]{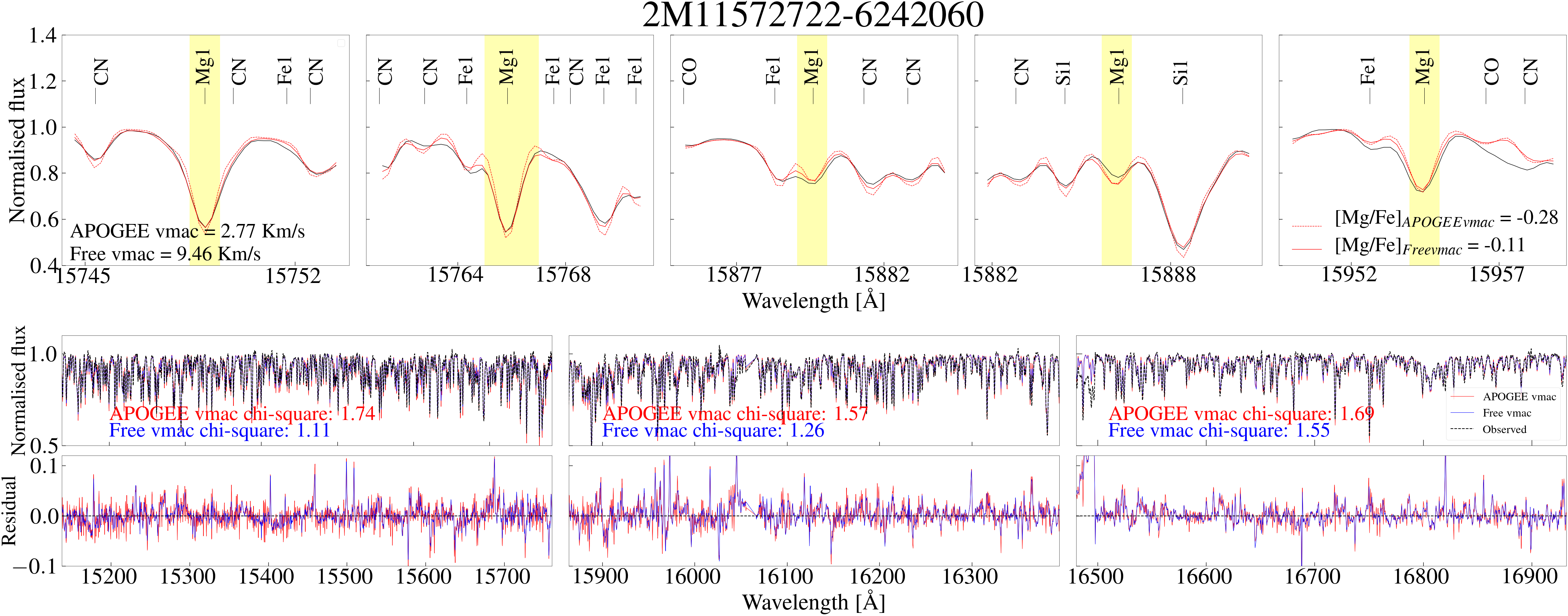}
  \caption{ Same as figure~\ref{fig:apogee_spectra_1}, for the massive sequence star 2M11572722-6242060. }
  \label{fig:apogee_spectra_3}
\end{figure*}

\begin{figure*}[]
 \centering
  \includegraphics[width=0.9\textwidth]{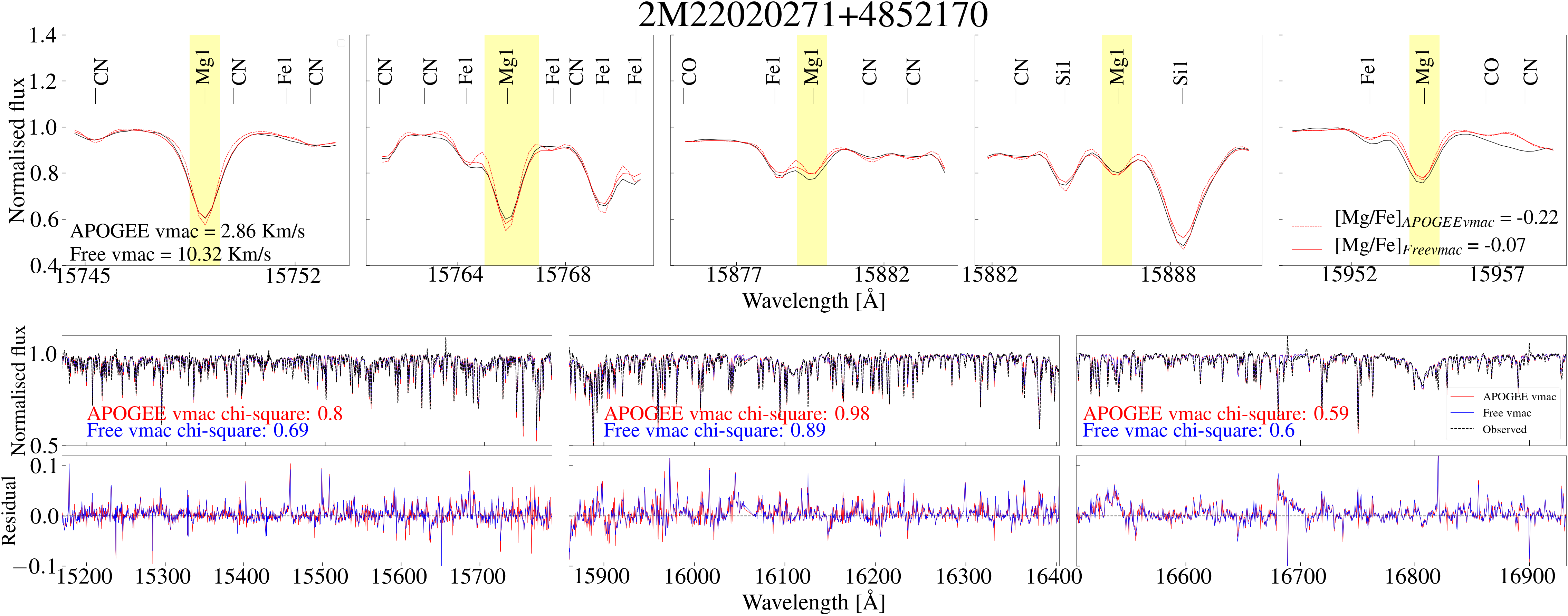}
  \caption{ Same as figure~\ref{fig:apogee_spectra_1}, for the massive sequence star 2M22020271+4852170. }
  \label{fig:apogee_spectra_4}
\end{figure*}

\end{appendix}

\end{document}